\documentclass[fleqn,usenatbib]{mnras}

\usepackage{newtxtext,newtxmath}
\usepackage[T1]{fontenc}

\DeclareRobustCommand{\VAN}[3]{#2}
\let\VANthebibliography\thebibliography
\def\thebibliography{\DeclareRobustCommand{\VAN}[3]{##3}\VANthebibliography}

\usepackage{graphicx}	% Including figure files
\usepackage{amsmath}	% Advanced maths commands
\usepackage{mathtools}
\title[Inward and outward migration]{Inward and outward migration of massive planets: moving towards a stalling radius}

\author[C. E. Scardoni et al.]{Chiara E. Scardoni,$^{1}$\thanks{E-mail: ces204@cam.ac.uk}
Cathie J. Clarke,$^{1}$
Giovanni P. Rosotti,$^{2,3}$
Richard A. Booth,$^{4}$\newauthor
Richard D. Alexander,$^{2}$
and Enrico Ragusa.$^{2,5}$
\\
$^{1}$Institute of Astronomy, University of Cambridge, Madingley Road, Cambridge CB3 0HA, UK\\
$^{2}$School of Physics and Astronomy, University of Leicester, University Road, Leicester LE1 7RH, UK\\
$^{3}$Leiden Observatory, Leiden University, PO Box 9513, 2300 RA Leiden, The Netherlands\\
$^{4}$Astrophysics Group, Imperial College London, Blackett Laboratory, Prince Consort Road, London SW7 2AZ, UK\\
$^{5}$Univ Lyon, Univ Lyon1, Ens de Lyon, CNRS, Centre de Recherche Astrophysique de Lyon UMR5574, F-69230 Saint-Genis-Laval, France
}

\date{Accepted XXX. Received YYY; in original form ZZZ}

\pubyear{2015}

\begin{document}
\label{firstpage}
\pagerange{\pageref{firstpage}--\pageref{lastpage}}
\maketitle

% Abstract of the paper
\begin{abstract}
Recent studies on the planet-dominated regime of Type II migration showed that, contrary to the conventional wisdom, massive planets can migrate outwards. Using `fixed-planet' simulations these studies found a correlation between the sign of the torques acting on the planet and the parameter $K'$ (which describes the depth of the gap carved by the planet in the disc).
We perform `live-planet' simulations exploring a range of $K'$ and disc mass values to test and extend these results. The excitation of planet eccentricity in live-planet simulations breaks the direct dependence of migration rate (rate of change of semi-major axis) on the torques imposed, an effect that `fixed-planet' simulations cannot treat. By disentangling the contribution to the torque due to the semi-major axis evolution from that due to the eccentricity evolution, we recover the relation between the magnitude and sign of migration and $K'$ and argue that this relation may be better expressed in terms of the related gap depth parameter $K$.  
We present a toy model in which the sign of planetary migration changes at a limiting value of $K$, through which we explore planets' migration in viscously evolving discs. The existence of the torque reversal shapes the planetary system's architecture by accumulating planets either at the stalling radius or in a band around it (defined by the interplay between the planet migration and the disc evolution). In either case, planets pile up in the area $1-10$ au, disfavouring hot Jupiter formation through Type II migration in the planet-dominated regime.
\end{abstract}

% Select between one and six entries from the list of approved keywords.
% Don't make up new ones.

%%%%%%%%%%%%%%%%%%%%%%%%%%%%%%%%%%%%%%%%%%%%%%%%%%

%%%%%%%%%%%%%%%%% BODY OF PAPER %%%%%%%%%%%%%%%%%%

\section{Introduction}
Planets form from protoplanetary disc material and continue to interact with this material by exchanging orbital energy and angular momentum via tidal torques. This interaction shapes both planetary orbital architecture and disc structure.
Planet-disc interactions have been studied over many decades, long before the detection of the first exoplanet by \citet{Mayor&Queloz1995}: see for example \citet{Lin&Papaloizou1979b,Goldreich&Tremaine1979,Goldreich&Tremaine1980}. Recent reviews of the topic include \citet{Papaloizou&Terquem2006,Kley&Nelson2012,Baruteau+2014,Papaloizou2021,Paardekooper+2022}.

Once a planet has formed in a protoplanetary disc, depending on its mass it might be able to open a gap in the disc surface density, or it may remain embedded in the disc. On this basis, planet migration can be classified \citep[e.g.][]{Artymowicz1993,Ward1998,Kley&Nelson2012} as type I migration \citep{Ida&Lin2008,Bitsch+2013}, for light planets that migrate embedded in the disc; or Type II migration, for massive planets that open a gap in the disc.
In this paper, we are interested in Type II migration \citep[e.g.][]{Lin&Papaloizou1993,Syer&Clarke1995,Ivanov+1999}, which can be further split into two regimes, defined through the local disc to planet mass ratio
\begin{equation}
    B_0=\frac{4\pi\Sigma_0 r^2}{m_{\rm p}}.
    \label{eq:B}
\end{equation}
The `disc-dominated' regime is defined as the regime where the local disc mass is higher than the planet mass; while the `planet-dominated' regime, conversely, requires the planet to be more massive than the local disc.

The classical theory of the disc-dominated regime assumes that the planet is locked in the disc gap, with no possible flux of material crossing its orbit, and it migrates following the disc viscous evolution \citep[e.g.][]{Lin&Papaloizou1979b}. In recent years, this simple picture has been questioned by hydrodynamics simulations' results; for example, \citet{Duffell+2014} and \citet{Duermann&Kley2015} suggested that planet migration is unlocked from the disc viscous evolution and the planet can migrate faster than the disc material; other works \citep[e.g.][]{Lubow&Dangelo2006} modelled the gas flow through the planetary gap. A number of recent studies investigated the problem \citep[e.g.][]{Robert+2018,Kanagawa+2018,Scardoni+2020,Lega+2021}, and \citet{Scardoni+2020} proposed that the hydrodynamic simulations mostly undergo a transient phase and over longer viscous timescales the planet migration conforms to the usual Type II picture; however, no definitive conclusion has been reached yet.

Strongly related to the migration problem, is the study of the gap's properties, since most of the angular momentum is exchanged between the planet and the disc in proximity of the gap edges \citep{Goldreich&Tremaine1979,Goldreich&Tremaine1980}; some important works analysing the gap shape and width are for example those by \citet{Crida+2006}, who provided an analytical formula linking the gap's and the planet-disc system properties; \citet{Fung+2014}, who analysed the expected density in the gap; and by \citet{Fung&Chiang2016}, who demonstrated that in 3D simulations, massive planets carve gaps whose properties are consistent with those produced in 2D simulations.

Regardless of the details of migration velocity in this regime, Type II migration is often suggested as an explanation for the existence of massive exoplanets characterised by small semi-major axes (the so-called `hot Jupiters'). Moreover Type II migration in the
planet-dominated regime -- the main focus of this paper -- might help in explaining the population of Jupiter-like planets located at larger radii. Indeed, in the latter regime, the planet's inertia is supposed to have a crucial impact on migration, slowing it down significantly. 

The very first papers exploring this problem \citep{Syer&Clarke1995,Ivanov+1999}, studied the problem in 1D under the assumption that no material can cross the planet's location. Since the planet migrates more slowly than the local viscous velocity and the material cannot cross the gap, the fate of the inner disc is to disappear, as it is rapidly accreted by the central star. For the same reason the gas located in the outer disc tends to accumulate at the outer gap edge. As the surface density of the gas at the outer gap edge increases, the torque pushing the planet inwards is enhanced, until reaching an equilibrium, where the planet moves at a rate matching the inward motion of the density maximum.

More recent studies have analysed the planet-dominated regime in different conditions, finding that in some circumstances the planet migration is not only slowed down, but even reversed.
\citet{Crida&Morbidelli2007}, for example, studied the problem in the presence of a gap with a non-negligible amount of material. Adding the corotational torque in their computations, they found that it can have a significant influence on planet migration: if the Reynolds number is low enough the planet migrates outwards instead of inwards towards the star. Under the assumption that the planet migration is locked to the viscous evolution, another situation for potential outward migration is the case that the planet is initially located in the disc area which is viscously expanding \citep{Veras&Armitage2004}. \citet{Hallam&Paardekooper2018} instead showed that, under certain circumstances, the illumination of the outer gap edge by the central star's radiation can lower the torque exerted by the outer disc on the planet; they thus suggested that this is a possible mechanism to  slow down or even reverse massive planet migration.

Further investigation on the planet migration direction has been recently conducted by \citet{Dempsey+2020,Dempsey+2021}, who used a set of 2D systems in the planet-dominated regime, in a stable steady state condition. To study planet migration they model the situation where the planet's orbital parameters (semi-major axis and eccentricity) are not allowed to evolve; they then calculate planet migration from the torques acting on the fixed planet. Since the planet cannot react to the disc either in $a_{\rm p}$ or in $e_{\rm p}$, the migration rate for given system parameters is just proportional to the disc mass; migration is parametrised in terms of $\Delta T/(\dot{M}l_{\rm p})$, where $\Delta T$ is the torque acting on the planet, $\dot{M}$ is the steady state accretion rate, $l_{\rm p}$ is the specific angular momentum of the planet. Through their study, they found that for typical disc parameters, Jupiter-like (or lighter) planets migrate inwards, while super-Jupiter planets migrate outwards; interestingly, the sign of the torque (and thus direction of migration) seems to be related to the gap parameter $K'=q^2/(\alpha h^3)$ (which controls the gap depth, \citealt{Kanagawa+2016}), where $q=m_{\rm p}/M_*$ is the planet to star mass ratio, $\alpha$ is the \citealt{Shakura&Sunyaev1973} viscosity parameter, and $h=H/r$ is the disc aspect ratio.
{The parameter $K'$ is a variant of the gap depth parameter $K=q^2/(\alpha h^5)$, which can be defined from the study of the torques' balance \citep[see for example][]{Fung+2014,Kanagawa+2015,Kanagawa+2018}. Being proportional to the planet's gravitational torque ($\propto q^2 / h^3$) and inversely proportional to  the disc's viscous torque ($\propto \nu \propto \alpha h^2$), it can be interpreted as a measure of the relative strength of the torques.}

These interesting results, however, are limited by the assumption of a fixed planet. Albeit in a different migration regime ($B>1$, see \equationautorefname~\ref{eq:B}), \citet{Scardoni+2020} showed that when the planet is allowed to change its orbital parameters, the disc density readjusts to the presence of the moving planet, which modifies the disc-planet interaction.
Although the two migration regimes cannot be directly compared, this work highlights the potential importance of including the evolution of the planet orbital parameters in the computation of disc-planet torques. The importance of the gap-adjustment due to planet migration in determining the evolution of the planet orbital parameters has recently also been underlined also by \citet{Lega+2021}. {Furthermore, in live planet simulations, the torque exerted on the planet by the disc modifies both the semi-major axis and the eccentricity \citep[e.g.][]{Kley&Dirksen2006,Duffell&Chiang2015,Teyssandier&Ogilvie2016,Rosotti+2017,Ragusa+2018}, potentially producing a non-trivial relation between the torque and planet migration.}

Another fundamental aspect of planet-disc interaction is the long timescale over which the disc re-adjusts itself to the presence of the planet; in fact, although the timescale of tidal interaction is short, the disc structure change might happen over the significantly longer viscous timescale \citep[e.g.][]{Ward1998, Lin&Papaloizou1979b}. The importance of considering the long term evolution of planet-disc systems has been demonstrated by \citet{Ragusa+2018}, who showed that the long term evolution obtained from hydrodynamic simulations can differ from the trend at the first stages of evolution. Apart from their work -- which was mainly focused on the eccentricity evolution -- no long term migration studies in the $B < 1$ regime have been conducted with a `live' (i.e. evolving) planet.

In this paper, we present a suite of 2D simulations lasting $\gtrsim 300$k orbits, for a variety of aspect ratios $H/r$, planet masses $M_p$ and disc masses $M_{\rm disc}$. These simulations are intended to be complementary to those by \citet{Dempsey+2021}, and allow us to investigate both the importance or otherwise of employing a `live' planet, as well as the role of different boundary conditions. Since these simulations are extremely expensive from the computational point of view, such simulations have never been performed before; consequently this is a new area of exploration for the disc-planet interaction problem, and the present simulation set inevitably leaves some questions unanswered. Nevertheless, they provide an intriguing starting point for further exploration in the field.
\begin{table*}
    \centering
    \caption{Simulation parameters. The name of each simulation is chosen in the following way: `L' or `M' to indicate a light or massive disc, respectively; `m' followed by a number to indicate the planet mass (measured in Jupiter masses); `h' followed by a number to indicate the aspect ratio.}
	\label{tab:simparameters}
    \begin{tabular}{lccccccccc}
        \hline
        {Name} & $B_0$ & $m_{\rm p}[m_{\rm J}]$ & $h_0$ & $r_{\rm in}$ & $r_{\rm out}$ & $\alpha_0$ & $K'_0$ & $K_0$ & $N_{\rm \rm orbits}$\\
        \hline
        L-m1-h036 & $0.046$ & $1$ & $0.036$ & $0.2$ & $15$ & $0.001$ & $21$ & $1.65 \cdot 10^4$ & $6\cdot 10^5$\\
        L-m3-h036 & $0.046$ & $3$ & $0.036$ & $0.2$ & $15$ & $0.001$ & $193$ & $1.49 \cdot 10^5$ & $3\cdot 10^5$\\
        L-m13-h036 & $0.046$ & $13$ & $0.036$ & $0.2$ & $15$ & $0.001$ & $3622$ & $2.9 \cdot 10^6$ & $3\cdot 10^5$\\
        L-m13-h06 & $0.046$ & $13$ & $0.06$ & $0.2$ & $15$ & $0.001$ & $782$ & $2.17 \cdot 10^5$ & $3\cdot 10^5$\\
        L-m13-h1 & $0.046$ & $13$ & $0.1$ & $0.2$ & $15$ & $0.001$ & $169$ & $1.69 \cdot 10^4$ & $3\cdot 10^5$\\
        L-m1-h06 & $0.046$ & $1$ & $0.06$ & $0.2$ & $15$ & $0.001$ & $5$ & $1.29 \cdot 10^3$ & $6\cdot 10^5$\\
        M-m1-h036 & $0.15$ & $1$ & $0.036$ & $0.2$ & $15$ & $0.001$ & $21$ & $1.65 \cdot 10^4$ & $6\cdot 10^5$\\
        M-m3-h036 & $0.15$ & $3$ & $0.036$ & $0.2$ & $15$ & $0.001$ & $193$ & $1.49 \cdot 10^5$ & $6\cdot 10^5$\\
        M-m13-h036 & $0.15$ & $13$ & $0.036$ & $0.2$ & $15$ & $0.001$ & $3622$ & $2.9 \cdot 10^6$ & $3\cdot 10^5$\\
        M-m13-h06 & $0.15$ & $13$ & $0.06$ & $0.2$ & $15$ & $0.001$ & $782$ & $2.17 \cdot 10^5$ & $3\cdot 10^5$\\
        M-m13-h1 & $0.15$ & $13$ & $0.1$ & $0.2$ & $15$ & $0.001$ & $169$ & $1.69 \cdot 10^4$ & $3\cdot 10^5$\\
        M-m1-h06 & $0.15$ & $1$ & $0.06$ & $0.2$ & $15$ & $0.001$& $5$ & $1.29 \cdot 10^3$ & $6\cdot 10^5$\\
        \hline
    \end{tabular}
\end{table*}

The paper is organised as follows: in \sectionautorefname~\ref{sec:Simulations} we describe the long numerical simulations performed to study planet-disc interaction; \sectionautorefname~\ref{sec:Results} contains the description of the planets' orbital parameters, and the analysis of the torques resulting from the disc to planet interaction; in \sectionautorefname~\ref{subsec:BCs} we discuss the influence of boundary conditions on the obtained torques on the planet; a toy model exploring the secular evolution of planets is developed in \sectionautorefname~\ref{sec:toymodel}, and in \sectionautorefname~\ref{sec:implications} we discuss the implications of this model for planetary demographics; finally, in \sectionautorefname~\ref{sec:Conclusions} we draw our conclusions.

\section{Simulations}
\label{sec:Simulations}
\subsection{Simulation parameters}
We performed 12 2D hydrodynamical simulations of protoplanetary discs containing a massive planet with the grid code {\sc Fargo 3D} \citep{BenitezLlambay&Masset2016}, considering a cylindrical reference frame $(r,\varphi)$ centered in the star. We adopt dimensionless units $G=M_*=r_0=1$, where $G$ is the gravitational constant, $M_*$ is the star mass and $r_0$ is the planet's initial location; the time unit is the inverse Keplerian frequency at $r_0$, $\Omega_{\rm k}^{-1}$; this means that the planet at its initial location $r_0$ requires $t=2\pi$ to complete an orbit. {Each simulation is run for $300$k orbits or $600$k, depending on the status of the steady state; the requirement is that $\dot{M}(r)$ is within 10\% of $\dot{M}(r_{\rm in})$ at least until $r=2.5\ a_{\rm p}$.}

The setup for the simulations is defined in analogy with the simulations by \citet{Ragusa+2018}.\footnote{In turn, their choice of parameters was based on the best fit for CI Tau disc by \citet{Rosotti+2017}} We consider logarithmically spaced $N_{r}=430$ cells in the radial direction, from $r_{\rm in}=0.2$ to $r_{\rm out}=15$, and linearly spaced $N_{\varphi}=580$ cells in the azimuthal direction, extending from $0$ to $2\pi$. We assume a locally isothermal equation of state set by \equationautorefname~\ref{eq:h}, and we parametrise the viscosity by applying the $\alpha$ prescription by \citet{Shakura&Sunyaev1973}, $\nu=\alpha c_{\rm s} H$, with $\alpha$ defined as a function of radius
\begin{equation}
    \alpha = \alpha_0 r^{-0.63};
\end{equation}
in all the simulations we fix $\alpha_0=0.001$.

In our simulations we consider a variety of disc aspect ratios and planet masses. Specifically, we consider flared discs, defining the aspect ratio as follows
\begin{equation}
    h=H/r=h_0 r^{0.215},
    \label{eq:h}
\end{equation}
where $h_0$ is the aspect ratio at the initial planet location, for which we consider 3 different values ($h_0=0.036,\ 0.6,\ 0.1$). Regarding the planet mass, we explore 3 different values: $m_{\rm p}=1\ m_{\rm J}$, $m_{\rm p}=3\ m_{\rm J}$, and $m_{\rm p}=13\ m_{\rm J}$.

For each combination of parameters $h_0$ and $m_{\rm p}$, we perform 2 simulations, characterised by different values for the initial local disc to planet mass ratio $B_0$ (see \equationautorefname~\ref{eq:B}).
The initial disc surface density at planet location $\Sigma_0$ is chosen to obtain the selected values for the initial value $B_0$; we thus have `massive' disc simulations, with $B_0=0.15$, and `light' disc simulations, with $B_0=0.046$. 
Note that in physical units, $B_0=0.046$ and $B_0=0.15$ for a Jovian mass planet at the radius of Jupiter corresponds to a local disc surface density of $1.3\ \rm g/cm^2$ and  $4.2\ \rm g/cm^2$, respectively. Apart from $m_{\rm p}$, $h_0$ and $\Sigma_0$, all the other parameters are kept fixed among all the simulations. {Being interested in the planet-dominated regime of Type II migration, we only considered values $B_0<1$; for planet migration with high B values we refer to our previous study \citep{Scardoni+2020}, specifically focused on the disc-dominated regime of Type II migration}

{In addition to the simulations outlined in \tableautorefname~\ref{tab:simparameters} we performed two high resolution simulations as convergence test: we performed simulations L-m13-h036 and M-m13-h036 with $\times 2$ and $\times 4$ the standard resolution used in this paper, and verified that the obtained orbital parameters are the same as those obtained using the standard resolution.}

\subsection{Initial and boundary conditions}
\label{subsec:Initial&BCs}
The initial density profile is defined as
\begin{equation}
    \Sigma(r) = \Sigma_0 r^{-0.3} \cdot {\rm e}^{\left(-{r}/{5}\right)^{1.7}},
\end{equation}
The planet mass is gradually increased during the first 50 orbits of the simulation, while its orbital parameters are kept fixed; once the planet reaches its full mass, it is allowed to evolve (i.e. to migrate and develop eccentricity) under the action of torques arising from planet-disc interaction.

We apply closed boundary conditions at the outer edge $r_{\rm out}=15$ by setting both the velocity and the gas density to zero at $r_{\rm out}$, i.e. no material is added to the simulation, in order to minimise the effect of the outer boundary condition. This generates a zero flux at $r_{\rm out}$, which physically corresponds to the case of a disc truncated by a wide binary.
At the inner boundary, instead, we apply `viscous' boundary conditions, i.e. we enforce the inner material velocity to match the viscous velocity (as in \citealt{Scardoni+2020} and \citealt{Dempsey+2020,Dempsey+2021}). This choice ensures that the inner boundary has a net zero effect on the total angular momentum budget of the disc, i.e. the torque supplied exactly matches the rate of angular momentum advected by accretion through the inner boundary.

In practice, the simulations are run for a small fraction of the viscous timescale at the outer edge so that the form of the outer boundary should not be critical to the calculations. {On the contrary, boundary effects can disturb the inner disc and thus the region around the planet on much shorter timescales, due to spurious wave propagation. We therefore employ the wave damping method by \citet{Val-Borro+2006} for the radial velocity at the inner boundary.} This means that over the damping timescale ($\tau=\Omega_{\rm k}/30$), we damp $v_r$ to the azimuthal average of the initial (viscous) velocity $\langle v_{r,0}\rangle$ in the region from $r_{\rm in}$ to $r=0.3$.

{We also perform an additional simulation characterised by the same parameters as simulation L-m13-h036, except for the inner radius and damping zone, which are taken to be half of the default values; thus we take $r_{\rm in}=0.1$, and the damping zone extends in this case up to $r=0.15$. This is to ensure that in the cases where the planet develops high eccentricity (and thus gets close to the inner boundary), the material in the damping zone does not affect the migration owing to numerical modification to the physical torque exerted on the planet.}

\section{Results}
\label{sec:Results}
\subsection{Orbital parameters}
\label{subsec:orbitalpar}
\begin{figure*}
    \centering
    \includegraphics[width=0.88\textwidth]{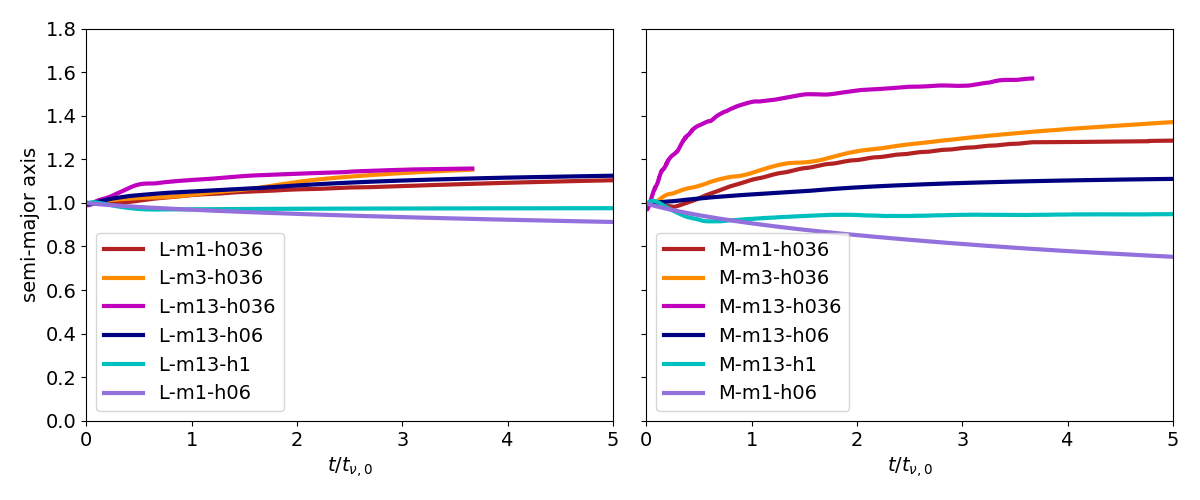}
    \caption{Semi-major axis as a function of the evolutionary time for light disc simulations (left panel) and massive disc simulations (right panel). The different colours correspond to different simulations, as indicated in the plot legend (see \tableautorefname~\ref{tab:simparameters} for the simulations' name definition).}
    \label{fig:migrationVISC}
\end{figure*}
\begin{figure*}
    \centering
    \includegraphics[width=0.88\textwidth]{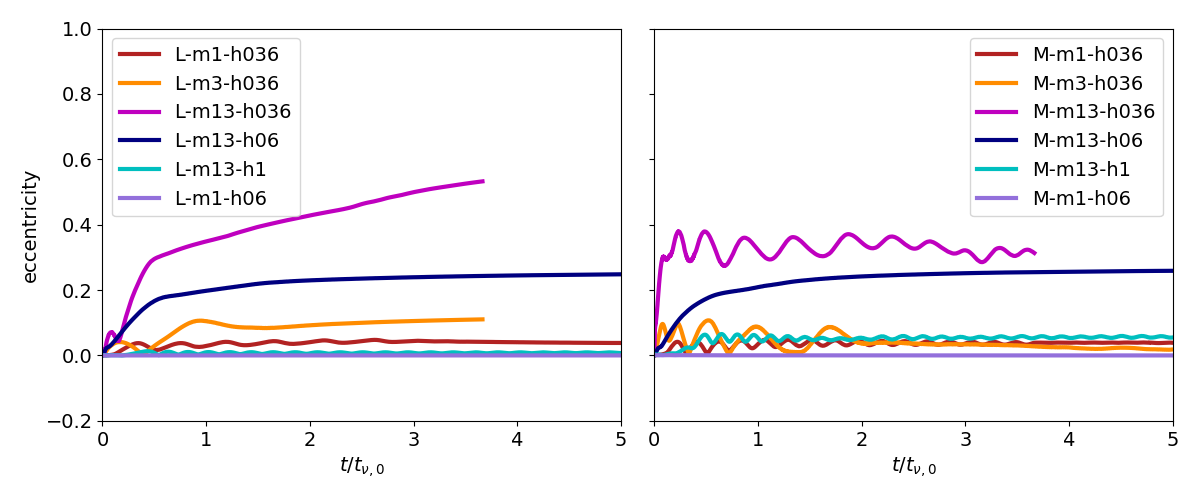}
    \caption{Eccentricity as a function of the evolutionary time for light disc simulations (left panel) and massive disc simulations (right panel). The different colours correspond to different simulations, as indicated in the plot legend (see \tableautorefname~\ref{tab:simparameters} for the simulations' name definition).}
    \label{fig:eccentricityVISC}
\end{figure*}
{In \figureautorefname~\ref{fig:migrationVISC} we show for the light disc simulations (left panel) and the massive disc simulations (right panel) the evolution of the semi-major axes as a function of evolutionary time, i.e. the physical time rescaled to the viscous timescale at the initial planet position
\begin{equation}
    t_{\nu}=\frac{2}{3}\frac{r^2}{\nu},
\end{equation} 
and we indicate with $t_{\nu,0}$ the viscous timescale evaluated at the initial planet location. For reference, the initial viscous timescales in our simulations are: $t_{\nu,0}\sim 9\cdot 10^5$ yr for $h=0.036$, $t_{\nu,0}\sim 3\cdot 10^5$ yr for $h=0.06$, $t_{\nu,0}\sim 10^5$ yr for $h=0.1$.} After an initial adjustment (lasting $t\lesssim t_{\nu,0}$), we can notice a variety of behaviours: inward migration (e.g. L-m1-h06, M-m1-h06), outward migration (e.g. L-m13-h036, M-m3-h036), or even stalling (e.g. L-m13-h1, M-m13-h06). 
This interesting behaviour can be related to the value of the gap-opening parameter $K'=q^2/(\alpha h^3)$ (where $q= m_{\rm p}/M_*$), whose analysis is deepened in \sectionautorefname~\ref{subsec:Torques}.

\figureautorefname~\ref{fig:eccentricityVISC}, instead, illustrates the planet eccentricity evolution. This plot shows a variety of behaviours, with some simulations (especially at higher planet masses) exhibiting significant eccentricity growth. It is also worth noticing that, in many cases, the planet eccentricity presents oscillations (as already observed, for example, by \citealt{Duffell&Chiang2015,Thun+2017,Rosotti+2017,Ragusa+2018}). These oscillations are due to the disc eccentricity vector evolving as a superposition of two rigidly precessing normal modes \citep{Teyssandier&Ogilvie2016,Teyssandier&Ogilvie2017,Ragusa+2018,Teyssandier&Lai2019}.

\subsection{Torques}
\label{subsec:Torques}
\subsubsection{Dependence on initial disc mass}
\label{subsec:dependenceoninitialdiscmass}
\begin{figure*}
    \centering
    \includegraphics[width=0.5\textwidth]{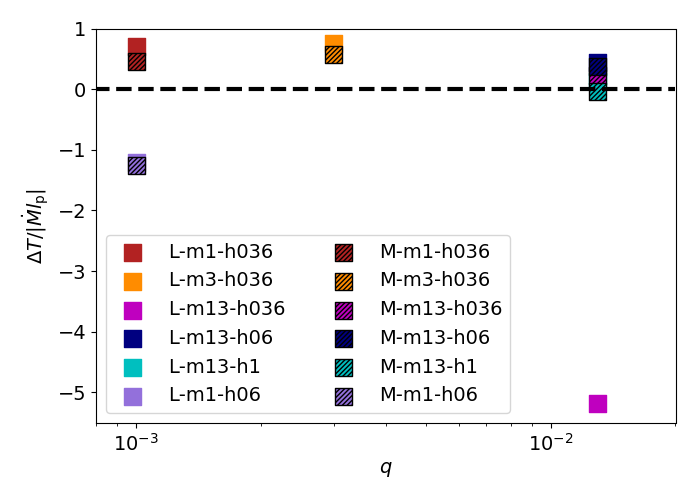}%
    \includegraphics[width=0.5\textwidth]{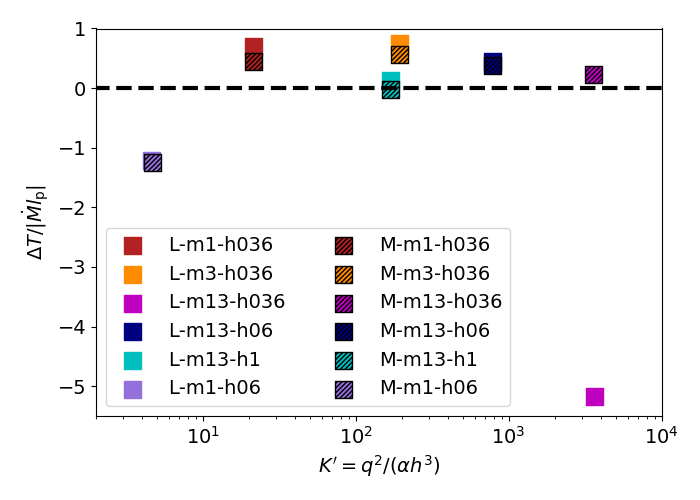}
    \caption{Torque acting on the planet normalised for the VSS accretion rate times the local specific angular momentum as a function of $q$ (left panel) and parameter $K'=q^2/(\alpha h^3)$ (right panel). The squares indicate the torque acting computed as variation rate of the planet's angular momentum obtained from the simulations; the different colours correspond to different simulations, as indicated in the legend.}
    \label{fig:JdotVSK}
\end{figure*}
In this section, we analyse the torque acting on the planet as a function of the following parameter \citep{Kanagawa+2015}
\begin{equation}
    K'=\frac{q^2}{\alpha h^3},
    \label{eq:parameterK'}
\end{equation}
which is a modification of the gap-opening parameter $K={q^2}/{(\alpha h^5)}$, i.e. a measure of the strength of the planet's gravitational torque ($\propto q^2 / h^3$) compared to the disc's viscous torque ($\propto \nu \propto \alpha h^2$).

The parameter $K'$ has been empirically studied by \citet{Kanagawa+2016,Kanagawa+2018} and by \citet{Dempsey+2020,Dempsey+2021}, who noticed that $K'$ correlates with the gap width and depth (with the gap becoming wider and deeper for higher $K'$ values). Systems characterised by the same $K'$ also appear to show similar torque density profiles; thus \citet{Dempsey+2020,Dempsey+2021} examined a number of fixed-planet\footnote{In fixed-planet simulations, the torque on the planet is computed as integration of the torque arising from the disc, from which planet migration is deduced; nonetheless, the planet orbital parameters are kept fixed at their initial values (i.e. $a_{\rm p}=1$, $e_{\rm p}=0$).} simulations characterised by different $K'$ values, and suggested that $K'$ is a good `ordering parameter' for planet migration. According to their analysis, in fact, planets in systems characterised by $K'\gtrsim 20$ migrate outward, whereas systems with $K'\lesssim 20$ present inward planet migration. Since their finding relies on simulations characterised by an evolving disc and fixed planets, our goal is to test whether the torque imposed on a fixed planet on a circular orbit is the same that applies in the case of a `live' planet which can respond to the torque during the evolution by changing its orbital elements.

For this purpose, we follow the approach of \citet{Dempsey+2021}, and plot the total torque on the planet normalised to $\dot{M}l_{\rm p}$ (see their Fig. 2). Note that for our convention of signs we have $\dot{M}<0$ for inward migration; to avoid confusion we therefore plot $\Delta T/|\dot{M}l_{\rm p}|$ whose sign only depends on the sign of the torque exerted by the disc on the planet (positive for outward migration and negative for inward migration).

The presence of a live planet, enables us to compute the torque acting on the planet directly as the planet's angular momentum variation rate
\begin{equation}
    \Delta T = \frac{{\rm d} J_{\rm p}}{{\rm d}t}=J_{\rm p} \left(\frac{1}{2a}\cdot\dot{a}-\frac{e}{{1-e^2}}\cdot\dot{e}\right),
    \label{eq:deltaT_ae}
\end{equation}
being $J_{\rm p}=m_{\rm p}\sqrt{GM_* a_{\rm p} (1-e_{\rm p}^2)}$.
Thus, we compute $\Delta T$ time-averaged over 30000 planet orbits,\footnote{ $30000$ orbits was chosen as being several times the timescale for periodic exchange of eccentricity between the planet and the disc, this being modulated on the beat period between the precession of the aligned and anti-aligned eccentric modes of the disc. } taken in the last stages of the simulation, when the inner disc has reached a steady state and the planet migration has stabilised. The results are shown in \figureautorefname~\ref{fig:JdotVSK}, from which we can notice that the general trend of positive torques for high $K'$, and negative torques for low $K'$ is confirmed,  {albeit with a suggestion of some decline in the torque at values of $K' > 20-200$}. Note, however, that the suite of simulations performed so far does not explore the $K'$ values extensively enough to confirm that $K'=20$ is the point of zero torque, and further simulations are needed for a precise characterisation of the $\Delta T - K'$ relation.

We further notice an applicability limitation in the results obtained from fixed planet simulations by \citet{Dempsey+2020,Dempsey+2021}. Those simulations do not conserve the total angular momentum of the system, because they do not allow the planet to evolve its orbital parameters. Although this approach is completely adequate to represent the limiting behaviour in the case that $B_0 \rightarrow 0$ (for which the planet is not expected to modify its orbital parameters noticeably over the disc lifetime), it cannot necessarily be used for $B_0\neq 0$ studies, where the planet might change its orbital parameters over the simulation timescale, and in the process affects the interaction with the disc.
This limitation becomes evident when we compare our results for different $B_0$ values (i.e. `light disc' vs `massive disc'), where we notice that the normalised torque properties change for different $B_0$ values in some cases. In the approach adopted by \citet{Dempsey+2021}, by contrast, both the torque and the accretion rate are assumed to vary linearly with $B_0$ and hence their ratio is independent of $B_0$; therefore in \citet{Dempsey+2020,Dempsey+2021} simulations the torque applied to the planet is, by construction, proportional to the disc mass. Our simulations demonstrate instead that although the assumption of torques independent of the disc mass is adequate for low mass planets, it breaks down in the case of high $K'$ ($>10^3$). 

We find that the normalised torque values are indeed approximately independent of $B_0$ in the case of the Jovian mass planet (see \figureautorefname~\ref{fig:JdotVSK}) but that for higher values of planet mass (and $K'$) the normalised torque values become increasingly divergent between the `light' and `massive' simulations. This is particularly marked in the case of the simulation with the largest $K'$ value (the $13\ m_{\rm J}$ planet with the lowest disc aspect ratio). These results can be readily understood from \figureautorefname~\ref{fig:eccentricityVISC} where it can be seen that the planet develops significant eccentricity during the course of the simulation. As discussed by \citet{Ragusa+2018,Teyssandier&Lai2019}, the eccentricity evolution of the planet depends on the interplay between eccentric modes of the disc whose structure depends on the disc to planet mass ratio. Thus we see in \figureautorefname~\ref{fig:eccentricityVISC} that not only do the simulation with largest $K'$ develop large eccentricities but that the eccentricity evolution is markedly different in the `light' and `massive' cases. It is therefore unsurprising that the torques on the planet evolve differently; indeed, in the light case, the eccentricity grows to the point where although the semi-major axis increases, the torque on the planet is actually negative (\figureautorefname~\ref{fig:JdotVSK}), a result that is reconciled by the large planetary eccentricity in this case. {We could thus consider simulation L-m13-h036 as an `outlier' in \figureautorefname~\ref{fig:JdotVSK} and \figureautorefname~\ref{fig:JdotVSKint} because it is the extreme case where the torque mainly affects $e_{\rm p}$ rather than $a_{\rm p}$.}

\begin{figure*}
    \centering
    \includegraphics[width=0.5\textwidth]{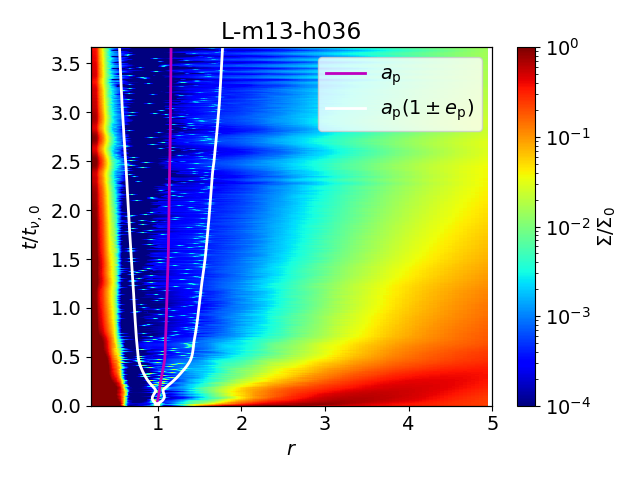}%
    \includegraphics[width=0.5\textwidth]{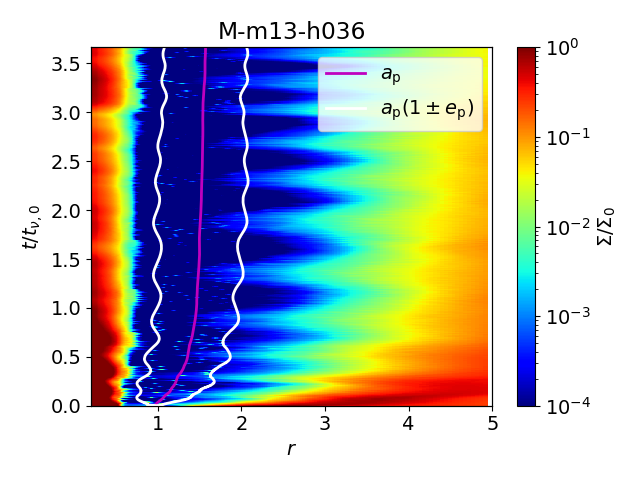}
    \caption{{Azimuthal averaged density profiles $\Sigma/\Sigma_0$ (colour map) as a function of the radius (x axis) and time normalised to the viscous timescale at the initial planet location (y axis). The magenta line shows the time evolution of the planet's semi-major axis; the white lines show the time evolution of $a_{\rm p}(1\pm e_{\rm p})$. The left panel refers to simulation L-m13-h036; the right panel refers to simulation M-m13-h036.}}
    \label{fig:densprofile_timeevolution}
\end{figure*}
{The different evolutionary histories of the disc structure in the light and massive 13 $m_{\rm j}$ case are illustrated in \figureautorefname~\ref{fig:densprofile_timeevolution} (in the left and right panel, respectively). The strong growth of planetary eccentricity in the light case is associated with the system settling into a single eigenmode of the system, i.e. the slow mode in which the apsidal precession of the disc and planet are aligned. In the massive case, conversely, the system exists in a state of superposition of aligned and anti-aligned modes \citep[see][for more discussion on the two modes]{Ragusa+2018}, resulting in the modulation of planetary eccentricity and disc structure evident in Figures 2 and \figureautorefname~\ref{fig:densprofile_timeevolution}.}\footnote{{In the light disc case with 13 $m_{\rm j}$ planet, the planet achieves such a high  eccentricity that the pericentre distance is only $\sim 2.5$ times the inner disc edge, $r_{\rm in}$. In order to check that the location of the inner boundary is not driving the evolution in this case, we have re-run the light disc case with $r_{\rm in}$ reduced by a factor 2 to 0.1. We show in \appendixautorefname~\ref{appendix1} that the evolution of the planetary orbital elements is unaffected by the reduction in $r_{\rm in}$.}}

We therefore conclude that the results based on non-migrating planet simulations presented by \citet{Dempsey+2020,Dempsey+2021} are a good approximation for light planets; nonetheless a more complex model, accounting for the planet's eccentricity growth and its dependence on disc mass, is needed when higher mass planets are considered.

\begin{figure}
    \centering
    \includegraphics[width=1\linewidth]{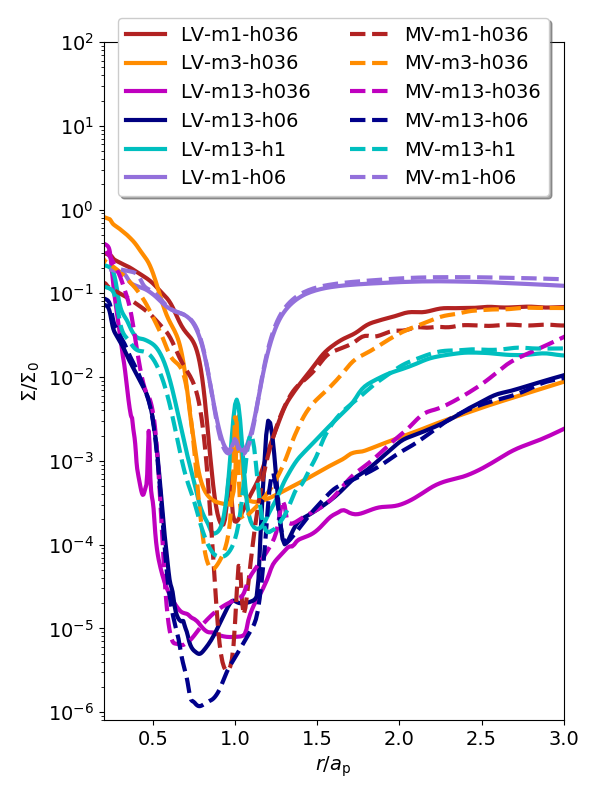}
    \caption{{Azimuthal averaged density profiles $\Sigma/\Sigma_0$ (colour map) at the end of the simulation time, plotted as a function of the radius normalised to the planet location {(both the density profile and the planet location are taken at the final snapshot of each simulation)}. Each colour refer to a different simulation, as indicated in the legend; the light and massive disc simulation with the same planet mass and disc aspect ratio are shown with the same colour with solid and dashed lines, respectively.}}
    \label{fig:densprofile_normalisedradius_allsims}
\end{figure}
{Finally we emphasise that \figureautorefname~\ref{fig:JdotVSK} is constructed after significant evolution of the disc profiles (i.e. after $6\cdot 10^5$ planetary orbits in the case of the massive and light simulations with $1\ m_{\rm j}$ planet, and in the massive disc simulation with $3\ m_{\rm j}$ planet; and after $3 \cdot 10^5$ planetary orbits in the remaining simulations) and use values of the torque and accretion averaged over the preceding $3 \cdot 10^4$ orbits. The result that the simulations with higher $K'$ values generally result in normalised torque values that are $\sim 0.5$ is all the more remarkable given the substantial variety in the surface density evolution that occurs in the various simulations. This is illustrated in \figureautorefname~\ref{fig:densprofile_normalisedradius_allsims} which shows snapshots of the surface density profiles (as a function of radius normalised to the current semi-major axis of the planet) in each of the simulations at the time that the normalised torques shown in \figureautorefname~\ref{fig:JdotVSK} are evaluated. In each case the solid and dashed versions of the curves relate to corresponding pairs of simulations with different values of $B_0$. The density is normalised to the initial surface density at the location of the planet in the unperturbed disc and therefore the relative surface density in each simulation pair can be obtained by scaling with the relevant values of $B_0$ for the light and massive disc simulations.}

{\figureautorefname~\ref{fig:densprofile_normalisedradius_allsims} demonstrates that it is only in the case of the simulation pair (L-m1-h06 and M-m1-h06)  with lowest $K'$ (i.e. narrowest gap) that the surface density profiles are simply scaled versions of each other. This can be traced to the fact that the planetary eccentricity is not excited in this case (see \figureautorefname~\ref{fig:eccentricityVISC}) as can be expected in the case of a narrow gap where corotation torques dominate over Lindblad resonances \citep{Goldreich&Tremaine1979,Goldreich&Tremaine1980}. At the other extreme, the simulations where the planetary eccentricity is most excited (i.e. the m13-h036 and m13-h1 pairs) show that the inner edge of the planet carved cavity is close to the 3:1 inner eccentric Lindblad resonance which is associated with strong driving of orbital eccentricity \citep{Lin&Papaloizou1993,Goldreich&Sari2003}.}

{We note that, in contrast to the simulations of \citet{Dempsey+2020,Dempsey+2021}, we do not drive the accretion rate at a prescribed rate in these simulations; the accretion rate at the inner edge (which is applied to the calculation of the normalised torque in \figureautorefname~\ref{fig:JdotVSK}) evolves rapidly during the first viscous timescale at the planetary radius, thereafter setting up a quasi-steady state within a few times the planetary radius (e.g. see \figureautorefname~\ref{fig:accretionprofile}). Thereafter, the magnitude of the accretion rate declines on the viscous timescale at the half mass radius of the disc, as expected. It is this slow decline in the accretion rate in the inner disc (and the associated normalisation of the surface density) that drives the leveling off in the evolution of the semi-major axis seen in \figureautorefname~\ref{fig:migrationVISC}. In \sectionautorefname~\ref{sec:toymodel} we will explore toy models which use the normalised torque values extracted from our simulations to calculate the secular evolution of planets in an evolving disc.}

\subsubsection{Torque components}
\begin{figure}
    \centering
    \includegraphics[width=1\linewidth]{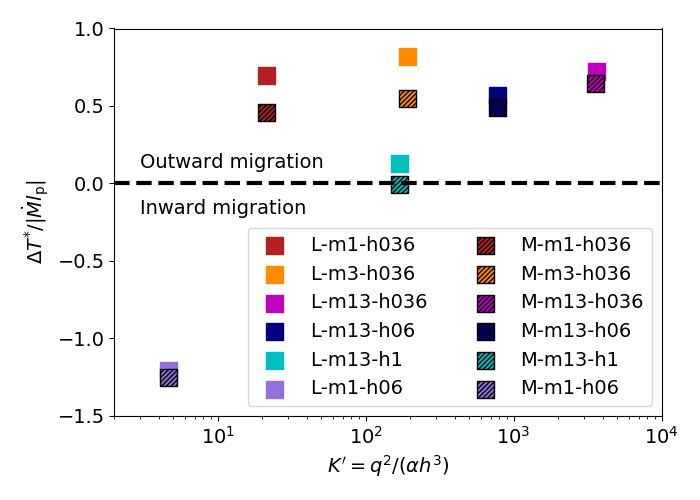}%
    \caption{{Torques associated to the rate of change of semi-major axis normalised for the VSS accretion rate times the local specific angular momentum as a function of parameter $K'=q^2/(\alpha h^3)$. The different colours correspond to different simulations, as indicated in the legend.}}
    \label{fig:JdotVSK_ecorrection}
\end{figure}
\begin{figure*}
    \centering
    \includegraphics[width=0.5\textwidth]{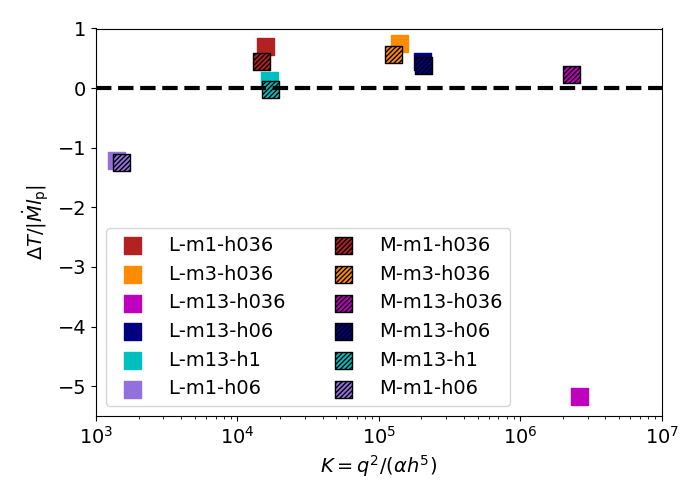}%
    \includegraphics[width=0.5\textwidth]{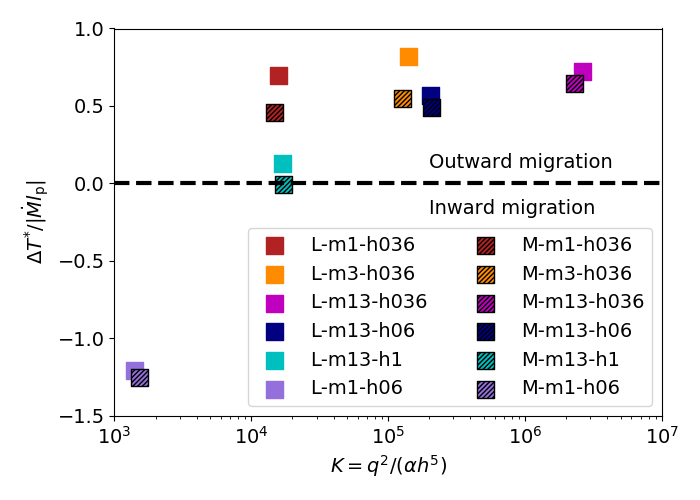}
    \caption{{Left panel: torque acting on the planet normalised for the VSS accretion rate times the local specific angular momentum as a function of $K$. Right panel: torques associated to the rate of change of semi-major axis normalised for the VSS accretion rate times the local specific angular momentum as a function of $K$. The different colours correspond to different simulations, as indicated in the legend.}}
    \label{fig:JdotVSKint}
\end{figure*}
So far we have evaluated the total torque on the planet, including contributions from both rate of change of semi-major axis and rate of change of eccentricity, as in \equationautorefname~\ref{eq:deltaT_ae}. In \figureautorefname~\ref{fig:JdotVSK_ecorrection} we plot instead the torque component associated with the rate of change of only the semi-major axis against the parameter $K'$
\begin{equation}
    \Delta T^* = \frac{m_{\rm p}}{2}\cdot \sqrt{\frac{GM_*}{a_{\rm p}}(1-e_{\rm p}^2)}\cdot \dot{a_{\rm p}}
    \label{eq:DeltaT*}
\end{equation}
We notice that in this case there is little difference between the torques associated to $\dot{a}_{\rm p}$ for the light and massive disc simulations with 13 $m_{\rm j}$ and $h=0.036$. This suggests that the differences in total torque between these two simulations do not result from interactions that transfer significant orbital energy to the planet, but are instead associated with interactions that primarily affect the planetary eccentricity. 

{We furthermore notice that at large $K'$ (or $K$, see the next section), the normalised $\Delta T^*$ is almost constant with $K'$ (or $K$), whereas $\Delta T$ seems to decrease. 
This behaviour suggests that as the planet mass increases, the resonances that excite/damp the planet's eccentricity are strengthened/ weakened; while the resonances responsible for migration seem less affected.}

This result has interesting implications on the rate of energy transfer to the planet. {Given that in this case the variation of normalised $\Delta T^*$ between simulations is much less than if the total normalised torque is considered, even the L-m13-h036 simulation (where the total normalised torque is large and negative: see \figureautorefname~\ref{fig:eccentricityVISC}) has a positive normalised $\Delta T^*$ value whose magnitude ($\sim ~ 0.5$) is similar to that in lower mass planets. This implies that the spread in normalised $\Delta T$ values for the most massive planets is driven by differences in eccentricity evolution which are not associated with significant contributions to the transfer of energy between disc and planet.} Given that
\begin{equation}
    E_{\rm p}=-\frac{GM_*m_{\rm p}}{2a_{\rm p}},
\end{equation}
the rate of change of planetary orbital energy is given by
\begin{equation}
    \dot{E}_{\rm p}=\frac{GM_*m_{\rm p}}{2a_{\rm p}^2}\dot{a}_{\rm p}.
    \label{eq:energy_variationrate}
\end{equation}
Combining \equationautorefname~\ref{eq:energy_variationrate} and the torque associated with the semi-major axis variation $|\Delta T^*|$ (\equationautorefname~\ref{eq:DeltaT*}) we deduce that
\begin{equation}
    \dot{E}_{\rm p}=\left(\frac{\Delta T^*}{\dot{M}l_{\rm p}}\right) \cdot \left(\frac{GM_*}{a_{\rm p}}\right) \cdot \dot{M}.
    \label{eq:Edot}
\end{equation}
{Since from our simulations the normalised $\Delta T^*$ is roughly constant (right panel in \figureautorefname~\ref{fig:JdotVSKint}), \equationautorefname~\ref{eq:Edot} implies that the energy transfer to the planet at a given location depends only on the accretion rate through the disc and is independent of planet mass.}

Moreover  we note that the migration rate, $\dot{a}_{\rm p}$, scales with $\dot{M}/m_{\rm p}$ and, for simulations with the same value of $B_0$ and $t_\nu$, should be independent of planet mass. This can be seen from the asymptotic gradients of the migration tracks shown in \figureautorefname~\ref{fig:migrationVISC}.

{ The analysis of the torque components allowed us to attribute the deviation of the torques $\Delta T$ from the constant value at high $K$ (or $K'$) to the planet eccentricity excitation, and we noticed that the effect becomes more and more important as $K$ ($K'$) increases. We plan to further explore the behaviour at high $K$ (or $K'$) in future works.}

\subsubsection{Choice of ordering parameter}
The torque analysis conducted so far is following \citet{Dempsey+2020,Dempsey+2021} in assuming that $K'$ is the appropriate ordering parameter. In \figureautorefname~\ref{fig:JdotVSKint} we instead plot the normalised torques as a function of 
\begin{equation}
    K=\frac{q^2}{\alpha h^5}
    \label{eq:parameterK}
\end{equation}
for all our simulations, considering both the total torque (left panel) and the torque associated with $\dot{a}_{\rm p}$ (right panel). While the number of simulations is too small to make a definitive judgement on the parameter that best captures the transition between narrow gaps with inward migration and broad gaps with outward migration, \figureautorefname~\ref{fig:JdotVSKint} suggests that this transition occurs at $K= 1.5 \cdot 10^4$. {We use this threshold in our `toy modeling'
presented in \sectionautorefname~\ref{sec:toymodel}}. Remarkably the normalised torque associated with $\dot{a}$ assumes a constant value of $\sim 0.5$ for higher values of $K$.

\section{Role of inner boundary conditions }
\label{subsec:BCs}
\begin{figure}
    \centering
    \includegraphics[width=1\linewidth]{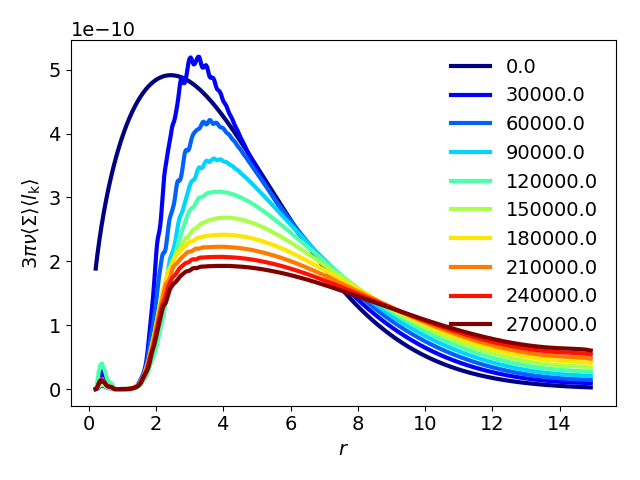}
    \caption{Viscous torque at different snapshots (showed in different colours, as indicated by the legend) for simulation M-m3-h036 with viscous boundary conditions.}
    \label{fig:ViscTorque}
\end{figure}

In this section, we discuss the role of boundary conditions in planet migration simulations.
For this purpose, we performed a set of additional simulations, characterised by the same parameters as those presented above, but with `open' boundary conditions at the inner boundary. The open boundary conditions allow the material to leave the grid at the inner edge at its own radial velocity (numerically, we set both the velocity and the density in the inner ghost cell equal to the last active cell). 
As a consequence, at $r_{\rm in}=0.2$ we have a zero torque boundary condition, as illustrated in \figureautorefname~\ref{fig:ViscTorque}, where the viscous torque is shown as a function of $r$ at different snapshots for simulation M-m3-h036 with open boundary conditions.\footnote{In this section, we use simulation M-m3-h036 as an example to illustrate the viscous torque and the accretion rate profile through the disc. Note, however, that the same behaviour is shown by all the other simulations in our sample.}

\begin{figure*}
    \centering
    \includegraphics[width=0.5\textwidth]{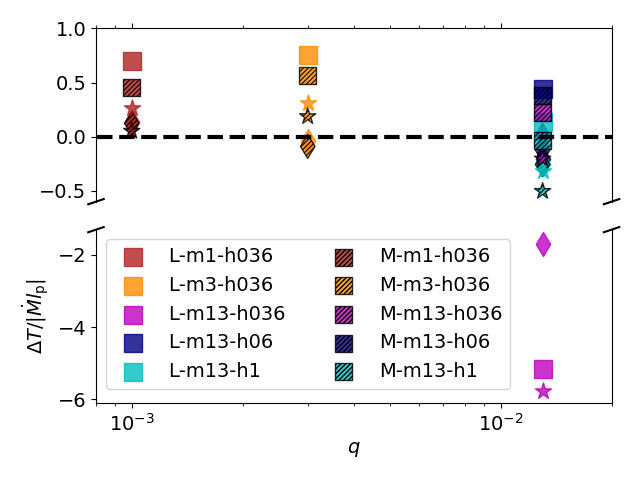}%
    \includegraphics[width=0.5\textwidth]{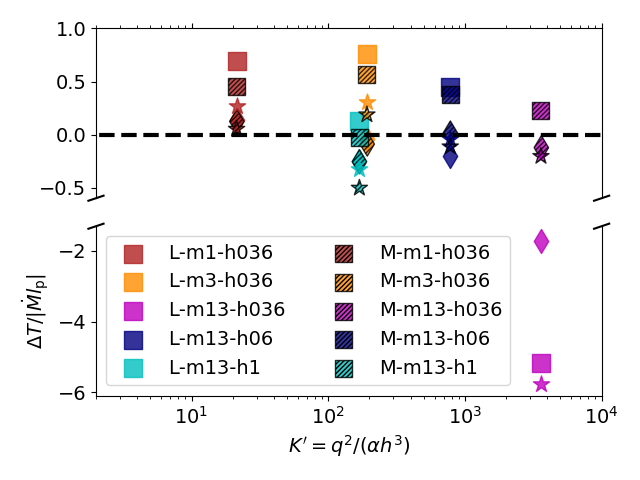}
    \caption{Torque acting on the planet normalised for the VSS accretion rate times the local specific angular momentum as a function of $q$ (left panel) and parameter $K'=q^2/(\alpha h^3)$ (right panel). The diamonds indicate the results from the open boundary condition simulations; the squares illustrate the torque the results from the viscous boundary condition simulations; the stars show the prediction for the open boundary condition results, applying \equationautorefname~\ref{eq:viscprediction} to the results obtained from viscous boundary conditions.}
    \label{fig:JdotVSK_openvisc}
\end{figure*}
From the physical point of view, the open boundary conditions produce a strong depletion of the inner disc; we can therefore apply them to study the physical case of a disc with a zero-torque inner cavity; for example, it may represent a case where the angular velocity is subject to a turning point (e.g. the classical boundary layer), or any case where material is removed from the cavity without injecting angular momentum to exterior material (e.g. photoevaporation or magnetospheric accretion).

\begin{figure}
    \centering
    \includegraphics[width=1\linewidth]{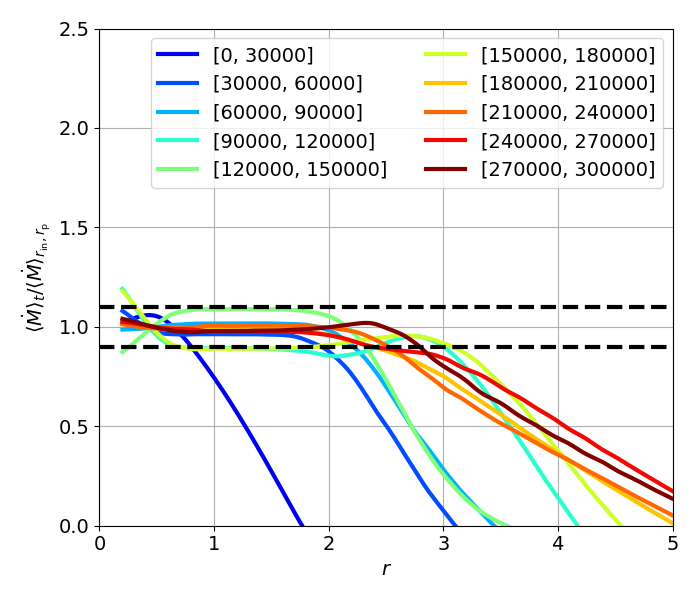}
    \caption{{Normalised accretion rate profiles time averaged over different time intervals (in the legend we indicate the initial and final orbit of the considered interval) for the massive disc simulation with $m_{\rm p}=3\ m_{\rm j}$, $h=0.036$ (M-m3-h036). The black dashed lines show the $10\%$ limits on the value of the normalised accretion rate, in order to consider the region in steady state.}}
    \label{fig:accretionprofile}
\end{figure}

The different physical configuration naturally produces differences in the migration properties which can be analysed in terms of $\Delta T^{\rm in}/|\dot{M}l_{\rm p}|$, where $\Delta T^{\rm in}$ is the torque imparted to the planet from the disc interior to the planet. In order to analyse the different torques acting on the planet in the two cases, we commence by noticing that for both the boundary condition choices, the inner disc (from $r_{\rm in}$ to a few $r_{\rm p}$) reaches a viscous steady state after some $t_{\nu,0}$ (see \figureautorefname~\ref{fig:accretionprofile}). If we consider the area of the disc between $r_{\rm in}$ and $r_{\rm p}$, the total angular momentum injected per unit time into this region is due to advection and viscous torques at $r_{\rm in}$ and $r_{\rm p}$ is:
\begin{equation}
    \Delta T^{\rm in}= -\dot{M} (l_{\rm p}-l_{\rm in})+F_{\nu}(r_{\rm in})-F_{\nu}(r_{\rm p})\sim -\dot{M} (l_{\rm p}-l_{\rm in})+F_{\nu}(r_{\rm in}),
\end{equation}
where the approximation is valid for planets massive enough to create a deep gap in the disc, so that $\Sigma(r_{\rm p})\sim 0$, and thus $F_{\nu}(r_{\rm p})=3\pi\nu\langle\Sigma\rangle\langle l_{\rm k}\rangle\sim 0$.
Since for open boundary conditions $F_{\nu}(r_{\rm in})\sim 0$, the torque in this case is
\begin{equation}
    \Delta T_{\rm open\ BC}^{\rm in}= -\dot{M} (l_{\rm p}-l_{\rm in});
\end{equation}
whereas for viscous boundary conditions $F_{\nu}(r_{\rm in})=-\dot{M} l_{\rm in}$, thus
\begin{equation}
    \Delta T_{\rm viscous\ BC}^{\rm in}= -\dot{M}l_{\rm p}.
\end{equation}

We can therefore estimate the difference in torque imparted by the inner disc
\begin{equation}
    \Delta T_{\rm open\ BC}^{\rm in}=\Delta T_{\rm viscous\ BC}^{\rm in}+\dot{M} l_{\rm in},
    \label{eq:viscprediction}
\end{equation}
where $\dot{M}<0$ (for our convention of signs), thus $\Delta T_{\rm viscous\ BC}>\Delta T_{\rm open\ BC}$.
Since in both cases, the quasi-steady state of the inner disc requires that $\Delta T$ is balanced by the torque applied by the planet, it follows that the form of the inner boundary condition should alone determine the normalised torque of the inner disc on the planet. The overall physical consequence is that planets in discs with an inner cavity are expected to be less prone to outward migration than planets in discs where the angular momentum lost by advection is balanced by the viscous torque of inward lying material.

Using \equationautorefname~\ref{eq:viscprediction} we can predict the open boundary conditions' results from the values obtained for viscous boundary conditions (and viceversa). We illustrate the extent to which this simple prediction of how the inner boundary condition affects the planet migration through the plot in \figureautorefname~\ref{fig:JdotVSK_openvisc}. In that plot, we show the torque acting on the planet (computed as time derivative of the planet angular momentum) for both the viscous boundary conditions (squares) and the open boundary conditions (diamonds). The stars, instead, show the prediction for the open boundary conditions' results, based on the viscous boundary conditions' results, using \equationautorefname~\ref{eq:viscprediction}, and assuming that the inner boundary condition does not affect the torque on the planet from the outer disc. Each colour corresponds to a different simulation, which is indicated in the legend.

{\figureautorefname~\ref{fig:JdotVSK_openvisc} shows that for the 1 Jupiter mass simulations with $h=0.036$, the difference in normalised torque is almost exactly that predicted by \equationautorefname~\ref{eq:viscprediction} (with $r_{\rm in} = 0.2$). Indeed, the outward migration is significantly slowed when a zero torque boundary condition is applied at $r_{\rm in} = 0.2$, because of the diminished torque delivered by the depleted inner disc.
{The prediction of \equationautorefname~\ref{eq:viscprediction} is  appropriate as long as (a) the inner disc is in steady state interior to the planet (as assumed in the derivation of \equationautorefname~\ref{eq:viscprediction}), and (b) the modification of the structure of the inner disc due to the changed boundary condition does not modify the torque exerted by the outer disc. This explains the quantitative agreement between the predictions of \equationautorefname~\ref{eq:viscprediction} and the change in total torque experienced by the planet shown in \figureautorefname~\ref{fig:JdotVSK_openvisc} for simulations characterised by low mass planets, where both the conditions are satisfied.
We have verified, through examination of radial profiles of accretion rate (see \figureautorefname~\ref{fig:accretionprofile}), that the condition (a) is satisfied in all simulations. We therefore attribute the fact that the quantitative agreement with \equationautorefname~\ref{eq:viscprediction} declines as the planet mass increases to the breakdown of assumption (b).}
This is particularly evident in the case of the light $13 m_{\rm j}$ simulation with $h=0.036$ where \equationautorefname~\ref{eq:viscprediction} fails to replicate the dependence of torque on boundary condition both in magnitude and in sign. It is notable that these parameters result in significant planetary eccentricity in the case of both boundary conditions but that the eccentricity growth is weaker in the case of the zero torque (cavity) boundary condition (compare \figureautorefname~\ref{fig:eccentricityVISC} with Figure 6 of \citealt{Ragusa+2018}).
The modification of the outer disc structure in response to the different orbital evolution of the planet means that the difference in total torque is not simply related to the effect of the boundary condition on the inner disc alone.}

\section{Exploration of secular evolution using a toy model}
\label{sec:toymodel}
In this section we define a toy model which provides a simplified prediction of massive planets' migration, under the following key assumption: (i) $\Delta T/\dot{M} l_{\rm p}$ is a function of $K$ (\equationautorefname~\ref{eq:parameterK}); (ii) $\Delta T/\dot{M} l_{\rm p}$ is independent of disc mass.\footnote{As underlined in \sectionautorefname~\ref{subsec:dependenceoninitialdiscmass}, $\Delta T/\dot{M} l_{\rm p}$ does not depend on the disc mass only when the planetary orbit maintains a low eccentricity; this is best satisfied for the $1 \rm m_{\rm J}$ case, which we therefore use in the toy model.} {Since the disc is flared, we expect $K\propto h^{-5}$ to vary with radius as $K\propto r^{-5 f}$, where $f$ is the flaring index; in this section we assume constant $\alpha$, but if we allowed it to vary with radius, it would have given a further contribution to the dependence of $K$ on $r$.}

We compute the planet migration time using the angular momentum definition $M_{\rm p}l_{\rm p}=M_{\rm p}\sqrt{GM_* r_{\rm p}}$\footnote{Here we use $r_{\rm p}$ instead of $a_{\rm p}$ because we are assuming circular orbits.}
\begin{equation}
    t_{\rm mig}=\frac{r_{\rm p}}{\dot{r}_{\rm p}} = \frac{1}{2}\cdot \frac{M_{\rm p} l_{\rm p}}{\Delta T},
\end{equation}
where $t_{\rm mig}<0$ ($>0$) means inward (outward) migration. Using the definitions $M_{\rm d}=4\pi r^2\Sigma$ and $\dot{M}=3\pi \nu\Sigma$, we can rewrite the migration timescale as
\begin{equation}
    t_{\rm mig}= \frac{1}{2}\cdot \left(\frac{|\dot{M}|l_{\rm p}}{\Delta T}\right) \cdot \left(\frac{M_{\rm p}}{M_{\rm d}}\right) \cdot \left(\frac{M_{\rm d}}{|\dot{M}|}\right).
\end{equation}
Noticing that $M_{\rm p}/M_{\rm d}=1/B$ and $M_{\rm d}/\dot{M}= t_{\nu}$\footnote{We define $t_{\nu}={4}{r^2}/3{\nu}$} this can be written as
\begin{equation}
    \frac{r_{\rm p}}{\dot{r}_{\rm p}} = \frac{1}{2}\cdot \left(\frac{|\dot{M}l_{\rm p}|}{\Delta T}\right)\cdot\left(\frac{t_{\nu}}{B}\right).
\end{equation}

We now consider 3 different regimes: 
\begin{itemize}
    \item[(a)] classical Type ii migration with $B \gg 1$, for which $r_{\rm p}/\dot{r}_{\rm p} = - t_{\nu}$, hence we deduce $\Delta T/\dot{M}l_{\rm p}=-0.5/B$ (see, for example \citealt{Lin&Papaloizou1979b,D'Angelo+2005,Scardoni+2020});
    \item[(b)] classical Type ii migration with $B \ll 1$, for which $r_{\rm p}/\dot{r}_{\rm p} = - t{\nu}/B$, hence we deduce $\Delta T/\dot{M}l_{\rm p}=-0.5$ (see, for example \citealt{Syer&Clarke1995,Ivanov+1999});
    \item[(c)] migration at high $K$ where $\Delta T/\dot{M}l_{\rm p}\sim 0.5$ from our numerical results.
\end{itemize}
We can combine the two limits (a) and (b) as $\Delta T/|\dot{M}l_{\rm p}|=-0.5/(B+1)$; then we put all of these considerations together to obtain
\begin{equation}
    \frac{r_{\rm p}}{\dot{r}_{\rm p}} =
    \begin{dcases}
        -\frac{B+1}{B}\cdot t_{\nu}\qquad & {\rm if}\  {K < K_{\rm lim}}\\
        \frac{1}{B}\cdot t_{\nu}\qquad & {\rm if}\ {K > K_{\rm lim}}
    \end{dcases}
    \label{eq:migcases}
\end{equation}
where {$K_{\rm lim}$} is the limiting value which determines the transition from inward to outward migration. For the purpose of the toy model, we take {$K_{\rm lim}=1.5\cdot 10^4$} {(as estimated from \figureautorefname~\ref{fig:JdotVSKint}).}

Given all these considerations, we can define a function {$g(K)$} such that\footnote{Note that we define {$g(K)$} in this way because we want it to be proportional to $\Delta T/|\dot{M}l_{\rm p}|$, which is the quantity that we have analysed in the previous sections.}
\begin{equation}
    \frac{r_{\rm p}}{\dot{r}_{\rm p}} = \frac{1}{{g(K)}}\frac{B+1}{B} t_{\nu}.
    \label{eq:newmigration}
\end{equation}
We therefore need a function with the following properties: for high {$K$} and low $B$, {$g(K)\sim 1$}, to obtain the lower case in \equationautorefname~\ref{eq:migcases}; for low {$K$} we want {$g(K)\sim -1$}, to recover the classical prediction for Type II migration (upper case in \equationautorefname~\ref{eq:migcases}). Note that in deriving the formula in \equationautorefname~\ref{eq:newmigration} we have not considered the case with high {$K$} and high $B$, because for the typical disc parameters and plausible planet location, it is unlikely to obtain this combination; {consequently, for high values of $B$ we expect the planet to migrate inward following the `classical' Type II migration rate (as analysed in \citealt{Scardoni+2020}).} We thus model the functional form of $g(K)$ as
\begin{equation}
    g(K)=\tanh \left(\frac{K-K_{\rm lim}}{W}\right),    
    \label{eq:g}
\end{equation}
where $W$ controls the width of the transition in {$K$} over which {$g(K)$} changes sign. We then estimate the planet migration by solving the following equation
\begin{equation}
    {\dot{r}_{\rm p}=r_{\rm p}g(K)\frac{B}{B+1} \frac{1}{t_{\nu}},}
    \label{eq:toymodel}
\end{equation}
where we underline that all the quantities on the right hand side are functions of $r_{\rm p}$ (and hence time).

Since planets are expected to migrate inward (outward) for {$K$} values smaller (bigger) than $K_{\rm lim}=1.5\cdot 10^4$, and because $K$ increases during inward migration and decreases during outward migration in the case of flaring discs ($f > 0$ in \equationautorefname~\ref{eq:r20} below), we expect them to always migrate towards the location of the disc where $K=K_{\rm lim}$. We then call {$r_{\rm lim}$} the radius corresponding to {$K_{\rm lim}=1.5\cdot 10^4$} which is therefore an `attractor' for migrating planets
\begin{equation}
    {r_{\rm lim}=\left(\frac{q^2}{K_{\rm lim}\alpha_0 h_0^5}\right)^{1/(a+5f)},}
    \label{eq:r20}
\end{equation}
where $\alpha=\alpha_0 r^{a}$ and $h=h_0 r^f$.

{ Note that in the toy model we do not include the effect of eccentricity growth, because we develop the toy model in the case of Jovian mass planets where eccentricity growth is very modest. We neither include the effect of the planet on the surface density profile, because by comparing the evolution of the disc with and without a Jupiter mass planet we found that the two systems behave similarly over the long timescales.}

\subsection{Toy model in a non-evolving disc}
\begin{figure}
    \centering
    \includegraphics[width=1\linewidth]{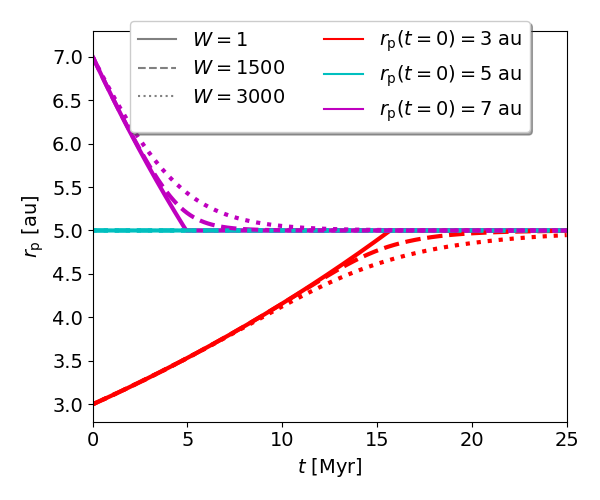}%
    \caption{Toy model for a planet of mass $m_{\rm p}=1\ m_{\rm j}$ and a disc characterised by $\alpha=0.001$, $h=0.02$ at $1\ \rm au$ and flaring index $0.25$, producing a stalling radius located at $5\ \rm au$. The different colours show the planet migration tracks for an initial planet location equal to 3 au (red lines), 5 au (cyan lines), and 7 au (magenta lines). The different line styles refer to different choices of parameter $W$: $W=1$ for the solid line; $W=1500$ for the dashed line; $W=3000$ for the dotted line. {At 5 au, the disc to planet mass ratio is $B=0.1$, while the viscous timescale is $t_{\nu}(5{\rm au})=1.6\ \rm Myr$}}
    \label{fig:toymodel}
\end{figure}
We first define a disc model characterised by $\alpha=0.001$ (constant throughout the disc), thus $\nu\propto r^{2f+1/2}$; and $\Sigma(r)=\Sigma(r_{\rm 1}) (r/r_{\rm 1})^{-1}$, where $r_1$ is the value of $1\ \rm au$ in code units, and $\Sigma(r_{\rm 1})$ is chosen to have $B_{5\rm au}=0.1$ at 5 au,\footnote{In the rest of the paper we will refer to the value of $B$ at 5 au as $B_{5\rm au}$; while we will indicate as $B$ the value at the planet location.} {corresponding to $M_{\rm disc}\sim 0.001\ \rm M_{\odot}$ for a disc of 100 au}. As the planet migrates (in either direction), the value of $B$ varies as\footnote{Note that the dependence of $B$ on $r_{\rm p}$ depends on the chosen density profile: if $\Sigma\propto r^{-s}$, then $B\propto r_{\rm p}^{2-s}$.}
\begin{equation}
    B(t) = B_{5\rm au} \cdot \left(\frac{r_{\rm p}(t)}{5\ \rm au}\right).
\end{equation}
To design a disc model with $r_{\rm lim}=5\ \rm au$ (i.e. the location of Jupiter), the disc aspect ratio at $1\ \rm au$ is taken equal to $0.02$, while the flaring index is $0.25$.

We then insert a Jupiter mass planet and we solve \equationautorefname~\ref{eq:toymodel}. In \figureautorefname~\ref{fig:toymodel} we illustrate the behaviour of {$r_{\rm lim}$} as an attractor, considering 3 different models: a planet initially located at $r_{\rm p}(t=0)=3\ \rm au\ <\ r_{\rm lim}$, which then migrates outwards until stalling at $5\ \rm au$ (red lines); for $r_{\rm p}(t=0)=5\ \rm au\ =\ r_{\rm lim}$ the planet stalls at its initial location (cyan line); for $r_{\rm p}(t=0)=7\ \rm au\ >\ r_{\rm lim}$, instead, the planet migrate inwards (magenta lines). The solid, dashed and dotted lines refer to values of parameter $W$ in \equationautorefname~\ref{eq:g} equal to $1$, $1500$, $3000$, respectively.

For the inward migrating planet (initially at $7\ \rm au$) the timescale required to reach the stalling radius in all cases is less than $10\ \rm Myr$, which means that the planet can reach the stalling location during the disc lifetime. In the case of the planet initially located at $3\ \rm au$, instead, the planet migrates to $4\ \rm au$ in $10\ \rm Myr$, but requires $\sim 15\ \rm Myr$ to reach the stalling radius. 
We caution, however, that the timescale on which the planet makes its final approach to the stalling radius is very sensitive to the choice of $W$ in the function $g(K)$; this underlines the importance of further simulations to explore the form of $g(K)$.
We furthermore notice that as the obtained migration timescale is comparable to the disc lifetime, we would expect the disc density to evolve in time over the planet's migration and model this possibility in the following section.

\subsection{Toy model in an evolving disc}
\label{sec:toymodel_evolvindisc}
\begin{figure*}
    \centering
    \includegraphics[width=0.5\linewidth]{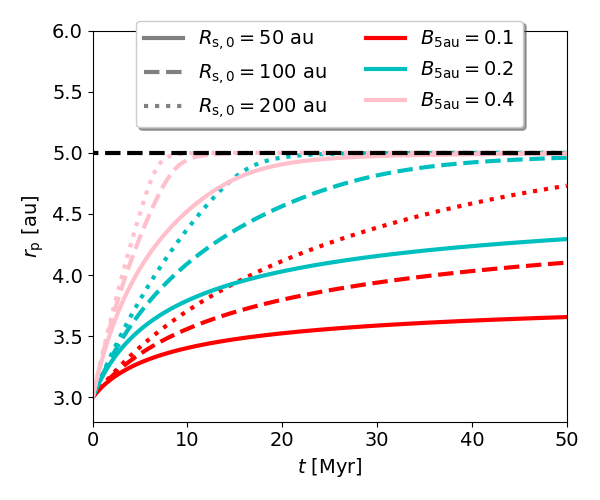}%
    \includegraphics[width=0.5\linewidth]{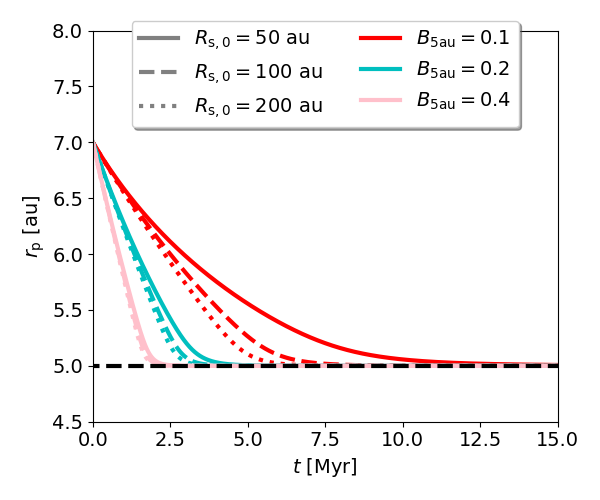}
    \caption{Toy model for a Jupiter mass planet intially located at $3$ au (left panel) and {$7$ au} (right panel), and migrating in a disc evolving according to the \citet{Lynden-Bell&Pringle1974} self-similar solution. The different colours and line styles refer to disc models characterised by a range of initial scale radii and $B_{5\rm au}$, whose values are indicated in the legend.}
    \label{fig:toymodel_evdens}
\end{figure*}
Since the planet migration timescale expected from the toy model is comparable to the disc lifetime, we define a variant of the toy model to take into account of the density evolution during migration.

For this purpose, while solving \equationautorefname~\ref{eq:toymodel} we evolve the density profile at each timestep according to the self-similar solution by \citet{Lynden-Bell&Pringle1974}
\begin{equation}
    \Sigma(R,t) = \frac{M_{\rm disc,0}}{2\pi R_{\rm s,0}}(2-b)\left(\frac{R}{R_{\rm s,0}}\right)^{-b}\tau^{-\eta}\exp\left[-\frac{(R/R_{\rm s,0})^{2-b}}{\tau}\right],
\end{equation}
where $M_{\rm disc,0}$ is the initial disc mass, defined by the choice of the density profile and the initial disc scale radius; $R_{\rm s,0}$ is the initial scale radius; $b$ is a parameter of the model which is the power law exponent for the
radial variation of the viscosity that we fix to 1 for consistency with the previous section; $\eta = (5/2-b)/(2-b)$; $\tau$ is defined as $\tau = 1+t/t_{\nu}$, with $t_{\nu}=R_{\rm s}^2/[3(2-b)^2\nu(R_{\rm s})]$. The surface density will thus evolve on the viscous timescale of the instantaneous scale radius
\begin{equation}
    R_{\rm s}(t)=R_{\rm s,0}\left[\frac{\tau}{2(2-b)}\right]^{\frac{1}{2-b}}.
\end{equation}

We show in \figureautorefname~\ref{fig:toymodel_evdens} the evolution tracks for a Jupiter mass planet initially located at $3\ \rm au$ (left panel) and $7\ \rm au$ (right panel), in a disc whose density is evolving as described above. In both cases, we consider different models, characterised by a range of initial scale radii (50 au with the solid lines, 100 au with the dashed lines, and 200 au with the dotted lines), and a range of $B_0$ (from 0.09 to 0.5, see the plot legend).
{For reference, the initial viscous timescale at the planet location are $t_{\nu}(3\ {\rm au})\sim 1\ \rm Myr$, and $t_{\nu}(7\ {\rm au})\sim 2.5\ \rm Myr$; while the viscous timescales at the chosen scale-radii are  $t_{\nu}(50\ {\rm au})=4.4\ \rm Myr$,  $t_{\nu}(100\ {\rm au})=8.9\ \rm Myr$,  $t_{\nu}(200\ {\rm au})=17.7\ \rm Myr$.}
Since the density is reduced while the disc evolves, the migration timescale becomes longer in both the cases when we consider an evolving disc (compare the red lines to \figureautorefname~\ref{fig:toymodel}). In fact, a decrease in the surface density causes the instantaneous value of the parameter $B$ to decrease, making the migration timescale longer. This means that models characterised by faster density evolution (i.e. those with lower $R_{\rm s,0}$) slow down the migration more effectively than those characterised by slower density evolution; this behaviour can be easily seen in \figureautorefname~\ref{fig:toymodel_evdens} by comparing the dotted, dashed, and solid lines with the same colour (i.e. with different $R_{\rm s,0}$ but same $B_{5\rm au}$).

Even more important is the effect of considering different $B_{5\rm au}$ values, with the migration timescale increasing while $B_{5\rm au}$ decreases. Furthermore we notice that in all the models the planet starting from $3$ au is more affected by the surface density time evolution, because the value of $B$ at its initial location is lower, thus it is more sensitive to further reductions in the local density value. In contrast, the planet starting from $7$ au reaches the stalling radius before decreasing significantly the value of $B$, so that it can reach the stalling radius in $\lesssim 10\ \rm Myr$ even in the model characterised by the lowest $B_{5\rm au}$ and the smallest $R_{\rm s,0}$ (see the red solid line in the right panel).

This result has interesting implication for the properties of a system characterised by outward migrating planets. From \figureautorefname~\ref{fig:toymodel_evdens}, we can deduce that to have a Jupiter mass planet migrating to $5\ \rm au$ in our model, we need to take either $B_{5\rm au}\gtrsim 0.4$, or we can decrease it to $B_{5\rm au}\gtrsim 0.2$ if we take a disc with $R_{\rm s,0}\gtrsim 100\ \rm au$. For lower $B_{5\rm au}$ values, even in the case of discs with large scale radii, the planet fails to migrate to the stalling radius in $\lesssim 10\ \rm Myr$; nonetheless, we still expect some outward migration for those planets, with a final radius which is determined by the disc lifetime rather than by the value of the stalling radius.

\subsection{Stalling radius and disc properties}
\begin{figure}
    \centering
    \includegraphics[width=1\linewidth]{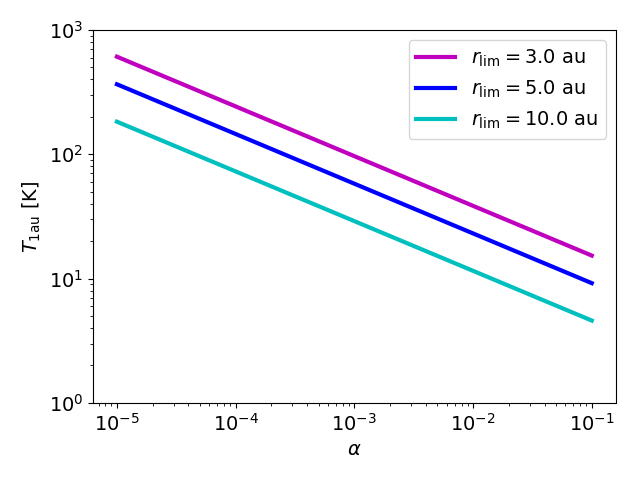}
    \caption{Disc temperature at 1 au as a function of the viscosity parameter $\alpha$, required to have a Jupiter mass planet stalling at $r_0$; the magenta line correspond to $r_0=3\ \rm au$, the blue line correspond to $r_0=5\ \rm au$, the cyan line correspond to $r_0=10\ \rm au$.}
    \label{fig:toymodel_Jup}
\end{figure}

In this section we focus on how the disc properties affect the location of the stalling radius. To investigate this problem we consider a Jupiter mass planet, and we compute the disc's properties required to have the planet stalling at a given radius $r_0$. In the left panel of \figureautorefname~\ref{fig:toymodel_Jup}, we consider 3 different stalling radii -- 3 au (magenta line), 5 au (blue line), 10 au (cyan line) -- and for each we plot the disc temperature $T$ at 1 au as a function of the $\alpha$ parameter. {Focusing first on the blue line, we notice that for a Jupiter mass planet that stalls 5 au the disc temperature at 1 au may vary from $T\gtrsim 10$ K to $T\gtrsim 10^2$ K, when we choose values for the $\alpha$ parameter in the range $\alpha\sim 10^{-4}-10^{-2}$.}\footnote{{Note that in our model we consider a constant $\alpha$ value, for which it is common to consider $\alpha\sim 10^{-4}-10^{-3}$; if we consider instead the fact that $\alpha$ depends on the specific location in the disc, direct turbulence measurements (e.g., \citealt{Carr+2004}) suggest that at sub-au radii $\alpha\sim 10^{-2}$ is more typical.}}
Since $T\gtrsim 10-10^2$ K at 1 au is a sensible temperature range for protoplanetary discs, this model is in agreement with the findings by \citet{Fernandes+2019} and \citet{Nielsen+2019} who suggest that around solar-like stars there is a peak in the number of giant planets located at $\sim 5$ au from the star. We further notice that for higher disc temperature the planet stalling radius decreases for fixed $\alpha$ because for a hotter disc, a larger region of the disc is in the low $K$ (shallow gap) regime where inward migration is expected.

If we consider more massive stars, the planet to star mass ratio decreases as $q\propto M_{*}^{-1}$, whereas the disc aspect ratio $h_0 \propto \sqrt{T/M_*}$, with $T\propto M_*^{\xi}$ (the relation is sub-linear, and the exact value of $\xi$ depends on the assumed stellar mass-luminosity relation). Using these scale relations in \equationautorefname~\ref{eq:r20} we find
\begin{equation}
    r_{\rm lim}\propto M_*^{\frac{1-5\xi}{2(a+5f)}},
\end{equation}
therefore we expect the stalling radius value to increase (decrease) with increasing stellar mass for $\xi <1/5$ ($\xi >1/5$). At fixed radius, the temperature approximately scales as $T\propto M_{*}^{0.15}$ \citep{Sinclair+2020}, we therefore expect that for the typical disc the location of the stalling radius increases for higher stellar mass; this is in agreement with the results from \citet{Nielsen+2019} that higher mass stars should have the peak of the Jupiter mass planet distribution at higher radius with respect to solar mass stars.

\section{Implications}
\label{sec:implications}
Planetary population synthesis models for giant planets  are widely based on the theory of Type II migration \citep{Ida&Lin2004a,Ida&Lin2008,Mordasini+2009a,Mordasini+2009b,Ida+2018,Bitsch+2019}. In its classic form this involves inward migration on a viscous timescale which decelerates at the point that the planet mass becomes comparable with the local disc mass. However, since the viscous timescale decreases with decreasing orbital radius, the net effect is that the planet arrives at the disc inner edge in a finite time \citep{Syer&Clarke1995,Ivanov+1999}. The continued inward driving by the disc is related to the assumption of zero flow past the planet so that material can always accumulate exterior to the planetary orbit and drive continued inward migration.

The present paper has demonstrated that even in calculations where the planetary orbital elements are allowed to evolve, giant planets do {\it not} arrive at the disc inner edge in a finite time. This is because as the planet migrates in to regions of the disc where the local disc mass dominates the planet mass, the disc establishes a quasi-steady state flow past the planet (cf \citealt{Dempsey+2020,Dempsey+2021}), preventing the inexorable extraction of angular momentum of the planet by material accumulating outside the planetary orbit. Moreover, as previously analysed by \citet{Dempsey+2020,Dempsey+2021} the sign of the torque depends on the quasi-steady state structure of the gas in the vicinity of the planet. A key point is that the disc interior to the planet is always in a steady state and so, in order to ensure net zero accumulation of angular momentum in this region, must impart a spin up torque to the planet of magnitude $\dot{M} l_{\rm p}$. On the other hand, the disc outside the planet is never globally in a steady state and thus the angular momentum that it extracts from the planet is not constrained to be $\dot{M} l_{\rm p}$, its magnitude  depending on the structure of the gap exterior to the planet. Thus for wider  gaps (either due to higher planet mass or lower disc aspect ratio $h$), the spin down torque on the planet is reduced as a result of the substantial clearing external to the planetary orbit: \figureautorefname~\ref{fig:densprofile_normalisedradius_allsims} demonstrates that the asymmetry between surface density interior and exterior to the planet becomes more pronounced for higher planet masses and colder (geometrically thinner) discs. We show that even in cases where the planet acquires a significant eccentricity as a result of interaction in the disc (i.e. those cases where the inner edge of the gap extends to close to the 3:1 mean motion resonance), the evolution of the planetary semi-major axis is well described in terms of a simple switch at a value of K (\equationautorefname~\ref{eq:g}) of around $1.5 \cdot 10^4$ (\figureautorefname~\ref{fig:JdotVSKint}).

We illustrate in \sectionautorefname~\ref{sec:toymodel} the implementation of some toy models that incorporate this evolutionary phenomenology. We examined the possibility that the location of zero torque might be imprinted on the architecture of planetary systems. As expected, the planetary evolutionary tracks demonstrate a convergence towards this attractor (\figureautorefname~\ref{fig:toymodel}), located at $5$ au given the planet mass and normalisation of the disc temperature profile (see \figureautorefname~\ref{fig:toymodel_Jup} for the sensitivity of the attractor location to disc parameters for a Jovian mass planet). However, the evolution may be too slow for planets to necessarily get close to the attractor location. First of all, the attainment of high $K$ values (associated with planet stalling) is favoured by relatively low alpha values where the viscous timescale is relatively long. Secondly, the speed of convergence upon the attractor location depends on the range of disc radii over which the torque magnitude undergoes a sign switch. Clearly, if this occurs relatively gradually, there will be a broad radial range where the torque values are low and therefore the evolution towards the attractor becomes very slow. Our present simulation set (right hand panel of \figureautorefname~\ref{fig:JdotVSKint}) does not allow us to tightly constrain the behaviour near the point of zero torque.
Finally, because the planet is not damming up the disc upstream of the planetary orbit, the disc's effect on the planet is progressively weakening on a timescale set by the viscous timescale of the outer disc. Thus whether the planet can be driven to the attractor depends on the migration timescale ($t_{\nu_0}/B$) compared with the viscous timescale of the outer disc: as illustrated in \figureautorefname~\ref{fig:toymodel_evdens}, efficient driving of the planet to its attractor location during the observed lifetime of protoplanetary discs \citep[e.g.][]{Alcala+2014,Manara+2016} is achieved for relatively high $B$ values (corresponding to fast migration timescale), and high $R_{\rm s,0}$ values (corresponding to slow viscous evolution).
Given these considerations, which depend on poorly constrained parameters such as the value of $\alpha$ and the form of the torque's dependence on gap shape in the region close to the location of zero net torque, it would be premature to argue that this process should impose a strong signature by piling up planets at a specific orbital location. Nevertheless, it is expected that Jovian mass planets should accumulate in the $1-10$ au range by this process.

What is certainly clear, however, is that these results preclude the production of hot Jupiters by disc mediated migration, at least in a non-self gravitating disc.{\footnote{Note that the contrary conclusion by \citet{Rosotti+2017} was a result of these authors not adopting viscous boundary conditions so that the loss of the inner disc led to a net negative torque on the planet: the \citet{Rosotti+2017} simulation corresponds to the diamond for the L-m13-h036 simulation in \figureautorefname~\ref{fig:JdotVSK_openvisc}}} This is because it is implausible that the disc gas surface density in the inner disc would ever be high enough to maintain $B>1$ (where roughly viscous timescale inward migration is expected) right down to the inner edge of the disc. For example, for a Jupiter mass planet at $0.1$ au, $B=1$ corresponds to such a high surface density that the total disc mass would likely exceed a solar mass within a few au. If the candidate hot Jupiter in the Classical T Tauri star CI Tau \citep{Johns-Krull+2016} is confirmed (see counterarguments by \citealt{Donati+2020}) it would then present a puzzle concerning how such a massive planet would have arrived in the innermost disc by an age of a few Myr. Rapid inward migration of giant planets during the earliest self-gravitating disc phase has been proposed by \citet{Baruteau&Paardekooper2011}, \citet{Malik+2015}, though more recent works suggest that even then it may not be possible to migrate into the innermost disc \citep{Stamatellos&Inutsuka2018,Rowther&Meru2020}.

{It is also interesting to notice that the existence of planet traps have already been suggested in a series of papers by \citet{Hasegawa&Pudritz2013,Hasegawa&Pudritz2014}, for lower mass planets (or planetary cores). They showed that low mass planets can be trapped at zero torque locations, potentially produced at disc radii characterised by significant density/thermal gradients, such as dead zone boundaries and ice lines.}

{We finally caution that this work relies on the assumption that only one planet is formed in the disc. In the case that multiple planets are formed, simulations including more migrating planets are needed.}

\section{Conclusions}
\label{sec:Conclusions}
In this work we have presented a suite of long term ($300-600\rm k$ planet orbits),  2D hydrodynamical simulations to test and expand recent findings by \citet{Dempsey+2020,Dempsey+2021} of an empirical correlation between the direction of planet migration and the value of the modified gap-opening parameter $K'$. We extend the torque analysis to systems with finite disc mass by considering  a live planet allowed to modify its orbital parameters.

Our live planet simulations confirm that there is a switch between inward and outward migration which is associated with the creation of deep gaps in the disc.
Gap depth is an increasing function of planet mass and a decreasing function of disc aspect ratio and can alternatively be parametrised by the quantities $K'=q^2/(\alpha h^3)$ (\equationautorefname~\ref{eq:parameterK'}) and $K=q^2/(\alpha h^5)$  (\equationautorefname~\ref{eq:parameterK}). Our simulations suggest that $K$ may be the better ordering parameter for describing planetary migration (see right hand panel of \figureautorefname~\ref{fig:JdotVSK_ecorrection}) and that the switch from inward to
outward migration occurs at $K_{\rm lim} = 1.5 \cdot 10^4$.

We further notice that, regardless the choice of the ordering parameter, as the planet mass increases some dependence on $B_0$ of the torques acting on the planet is revealed; this effect could not be seen in fixed planet simulations, where the normalised torque $\Delta T/|\dot{M}l_{\rm p}|$ values are disc mass independent by construction. {This means that if we consider low mass planets, the fixed-planet simulations by \citet{Dempsey+2020,Dempsey+2021} are a good approximation of the planets' migration; if we consider higher mass planets, the results are modified by the growth of the planet's eccentricity in a way that depends on the disc mass \citep{Ragusa+2018,Teyssandier&Lai2019}.}
Nonetheless we showed that, by disentangling the contribution to the torque due to the semi-major axis variation from the contribution due to the eccentricity evolution, the massive planet migration is well described by the change of sign of the disentangled torque at $K_{\rm lim}\sim 1.5 \cdot 10^4$.
It is a general feature of protoplanetary discs with realistic heating that the aspect ratio of the disc increases with radius and thus that $K$ is a decreasing function of radius. We thus predict inward (outward) migration for radii exterior (interior) to the location where $K$ has its limiting value.

We then model the migration behaviour of massive planets by describing the dependence of migration on the parameter $K$. We evaluate this migration in the context of the secular evolution of a viscous disc.
This model allows us to obtain the following results:
\begin{itemize}
    \item[(1)] since planets migrate inwards (outwards) for $K<K_{\rm lim}$ ($K>K_{\rm lim}$), they tend to go towards the location in the disc where $K=K_{\rm lim}$. We thus propose the existence of a `stalling radius' defined by the location where $K=K_{\rm lim}$ (i.e. the location of zero torque on the planet);
    \item[(2)] we study the dependence of the stalling radius on the disc parameters (temperature and $\alpha$ viscosity parameter), finding that typical disc parameters enable stalling radii in the range 3-10 au, in agreement with the peak in the Jupiter distribution at a few au \citep{Fernandes+2019,Nielsen+2019};
    \item[(3)] when we include the effect of disc density evolution in the model, the migration is slowed down, due to the density reduction with time (which reduces the parameter $B$). As a consequence, the planet migration towards the stalling radius might be limited by the disc lifetime in rapidly evolving systems characterised by relatively low $B_0$.
\end{itemize}
The toy model suggests that while planets should migrate towards a stalling radius set by the planet mass and disc aspect ratio, whether or not they attain their stalling radii depends both on the initial location of the planet and the over-all radial extent of the disc, since the latter determines the rate at which the disc surface density declines. Thus it is likely that this effect does not imprint a strong pile up of planets at their respective stalling radii but rather causes them to occupy a broad band of radii in the range $3-10$ au. Future quantification of the torque dependence on $K$ in the vicinity of $K_{\rm lim}$ will help to constrain this further.
In any case we do not expect planets to be able to move in from this band by disc mediated migration, thus posing a difficulty for hot Jupiter formation at early times.

\section*{Acknowledgements}
We would like to thank the referee for the careful reading of this paper and for the useful comments.
CES thanks Peterhouse Cambridge for a Ph.D. studentship. 
RAB acknowledges the Royal Society University Research Fellowship. GR acknowledges support from the Netherlands Organisation for Scientific Research (NWO, program number 016.Veni.192.233) and from an STFC Ernest Rutherford Fellowship (grant number ST/T003855/1). RDA and ER gratefully acknowledges funding from the European Research Council (ERC) under the European Union’s Horizon 2020 research and innovation programme (grant agreement no. 681601); RDA also acknowledges funding from STFC Consolidated Grant ST/W000857/1.
ER acknowledges funding from the European Research Council (ERC) under the European Union’s Horizon 2020 research and innovation programme (grant agreement No 864965).
This work has also been supported by the European Union’s Horizon 2020 research and innovation programme under the Marie Sklodowska-Curie grant agreement number 823823 (DUSTBUSTERS). 
This work was in part performed using the Cambridge Service for Data Driven Discovery (CSD3), part of which is operated by the University of Cambridge Research Computing on behalf of the STFC DiRAC HPC Facility (www.dirac.ac.uk). The DiRAC component of CSD3 was funded by BEIS capital funding via STFC capital grants ST/P002307/1 and ST/R002452/1 and STFC operations grant ST/R00689X/1.
This work was in part performed using the DiRAC Data Intensive service at Leicester, operated by the University of Leicester IT Services, which forms part of the STFC DiRAC HPC Facility (www.dirac.ac.uk). The equipment was funded by BEIS capital funding via STFC capital grants ST/K000373/1 and ST/R002363/1 and STFC DiRAC Operations grant ST/R001014/1. DiRAC is part of the National e-Infrastructure.

%%%%%%%%%%%%%%%%%%%%%%%%%%%%%%%%%%%%%%%%%%%%%%%%%%
\section*{Data Availability}
The code used to perform the simulations contained in this paper ({\sc Fargo3D}, \citealt{BenitezLlambay&Masset2016}) is available at \url{http://fargo.in2p3.fr}.

%%%%%%%%%%%%%%%%%%%% REFERENCES %%%%%%%%%%%%%%%%%%

% The best way to enter references is to use BibTeX:

\bibliographystyle{mnras}
\bibliography{biblio} % if your bibtex file is called example.bib

% Alternatively you could enter them by hand, like this:
% This method is tedious and prone to error if you have lots of references
%\begin{thebibliography}{99}
%\bibitem[\protect\citeauthoryear{Author}{2012}]{Author2012}
%Author A.~N., 2013, Journal of Improbable Astronomy, 1, 1
%\bibitem[\protect\citeauthoryear{Others}{2013}]{Others2013}
%Others S., 2012, Journal of Interesting Stuff, 17, 198
%\end{thebibliography}

%%%%%%%%%%%%%%%%%%%%%%%%%%%%%%%%%%%%%%%%%%%%%%%%%%

%%%%%%%%%%%%%%%%% APPENDICES %%%%%%%%%%%%%%%%%%%%%

\appendix
\section{Test inner boundary effects}
\label{appendix1}
\begin{figure*}
    \centering
    \includegraphics[width=0.5\linewidth]{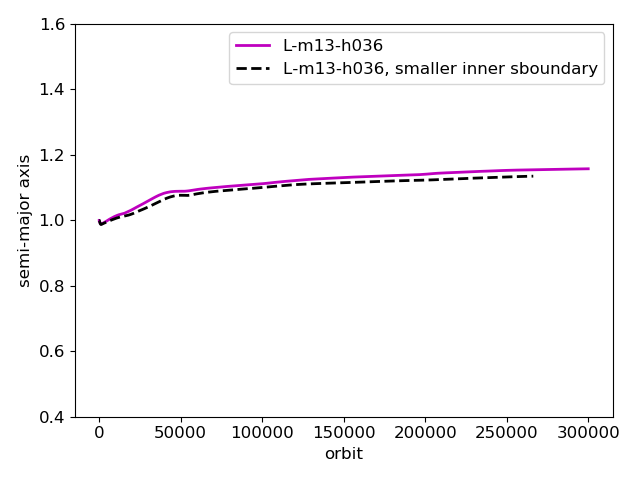}%
    \includegraphics[width=0.5\linewidth]{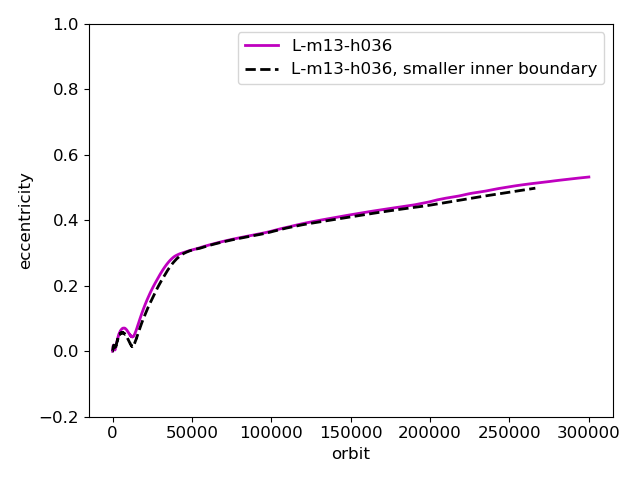}
    \caption{{Semi-major axis (left panel) and eccentricity (right panel) evolution in the light disc case with $13\ m_{\rm j}$ planet and $h=0.036$. The magenta line shows the planet evolution in the original simulation run L-m13-h=0.026, with $r_{\rm in}=0.2$; the black dashed line shows the result from the simulation test, characterised by $r_{\rm in}=0.1$.}}
    \label{fig:ae_testrin}
\end{figure*}
{From \figureautorefname~\ref{fig:eccentricityVISC} we noticed that in simulation L-m13-h036 (i.e. the light disc case with $13\ m_{\rm j}$ planet and $h=0.036$), the planet grows significantly its eccentricity so that its pericentre distance is only $\sim 2.5$ times the inner disc edge. Since at the inner boundary we are applying wave killing boundary conditions (see \sectionautorefname~\ref{sec:Simulations}), we must ensure that at the in simulation L-m13-h036 the planet evolution is not affected by numerical effects due to its closeness to the inner boundary when it is at the pericentre. We thus present here the results from a simulation test, characterised by all the same characteristics as those of simulation L-m13-h036, apart from the inner radius and the damping region, which are both taken to be half of their original values: we take $r_{\rm in}=0.1$, the damping is taken from $r_{\rm in}$ to $r=0.15$; we also increase the number of radial cells in order to obtain the same radial resolution as the original simulation run.}

{In \figureautorefname~\ref{fig:ae_testrin} we show the semi-major axis (left panel) and eccentricity (right panel) as a function of orbit, for both the original L-m13-h036 run (magenta line) and the simulation test with $r_{\rm in}=0.1$ (black dashed line). We notice that the semi-major and eccentricity evolution is essentially the same in both the runs, confirming that the planet evolution in simulation L-m13-h036 is not affected by boundary numerical effects.}

%%%%%%%%%%%%%%%%%%%%%%%%%%%%%%%%%%%%%%%%%%%%%%%%%%

% Don't change these lines
\bsp	% typesetting comment
\label{lastpage}
\end{document}